\documentclass[modern]{aastex631}
\usepackage[section]{placeins}
\usepackage{bm}
\newcommand{\Rmnum}[1]{\uppercase\expandafter{\romannumeral #1}}
\newcommand{\vect}[1]{\mathbf{#1}}
\newcommand{\B}[1]{{\color{blue}{#1}}}
\newcommand{\R}[1]{{\color{red}{#1}}}
\shorttitle{Current}
\shortauthors{Wang et al.}

\begin{document}

\title{Investigating pre-eruptive magnetic properties at the footprints of erupting magnetic flux ropes}

\correspondingauthor{Wensi Wang}
\email{minesnow@ustc.edu.cn}

\author{Wensi Wang}
\affil{CAS Key Laboratory of Geospace Environment, Department of Geophysics and Planetary Sciences\\
University of Science and Technology of China, Hefei, 230026, China}

\author{Jiong Qiu}
\affiliation{Department of Physic, Montana State University, Bozeman, MT 59717, USA}

\author{Rui Liu}
\affil{CAS Key Laboratory of Geospace Environment, Department of Geophysics and Planetary Sciences\\
University of Science and Technology of China, Hefei, 230026, China}
\affil{CAS Center for Excellence in Comparative Planetology, University of Science and Technology of China\\
Hefei 230026, China}
\affil{Mengcheng National Geophysical Observatory, University of Science and Technology of China\\
Mengcheng 233500, China}

\author{Chunming Zhu}
\affiliation{Department of Physic, Montana State University, Bozeman, MT 59717, USA}

\author{Kai E Yang}
\affiliation{Institute for Astronomy, University of Hawaii at Manoa, Pukalani 96768, USA}

\author{Qiang Hu}
\affil{Department of Space Science and CSPAR, University of Alabama in Huntsville, Huntsville, AL 35805, USA}

\author{Yuming Wang}
\affil{CAS Key Laboratory of Geospace Environment, Department of Geophysics and Planetary Sciences\\
University of Science and Technology of China, Hefei, 230026, China}
\affil{CAS Center for Excellence in Comparative Planetology, University of Science and Technology of China\\
Hefei 230026, China}

\begin{abstract}
It is well established that solar eruptions are powered by free magnetic energy stored in current-carrying magnetic field in the corona. It has also been generally accepted that magnetic flux ropes (MFRs) are a critical component of many coronal mass ejections (CMEs). What remains controversial is whether MFRs are present well before the eruption. Our aim is to identify progenitors of MFRs, and investigate pre-eruptive magnetic properties associated with these progenitors. Here we analyze 28 MFRs erupting within 45 deg from the disk center from 2010 to 2015. All MFRs' feet are well identified by conjugate coronal dimmings. We then calculate magnetic properties at the feet of the MFRs, prior to their eruptions, using Helioseismic and Magnetic Imager (HMI) vector magnetograms. Our results show that only 8 erupting MFRs are associated with significant non-neutralized electric currents, 4 of which also exhibit pre-eruptive dimmings at the foot-prints. Twist and current distributions are asymmetric at the two feet of these MFRs. The presence of pre-eruption dimmings associated with non-neutralized currents suggests the pre-existing MFRs. Furthermore, evolution of conjugate dimmings and electric currents within the foot-prints can provide clues about the internal structure of MFRs and their formation mechanism.  
\end{abstract}


\section{Introduction} \label{sec:intro}
Solar eruptions, such as prominence eruptions, solar flares and coronal mass ejections (CMEs), are the dominant contributor to adverse space weather at Earth. Now it is well established that these phenomena are powered by free magnetic energy, which is stored in current-carrying magnetic fields in the corona. What remains controversial is whether electric current is neutralized or not in the solar active regions (ARs) when integrating over the whole active region for each polarity individually. ARs are believed to be formed through subsurface flux tubes emerged from the solar interior \citep{fan2009emergence}. The current flowing in an isolated magnetic flux tube can be divided into two parts: the so-called main (direct) currents, and shielding (return) currents \citep{melrose1991neutralized,parker1996inferring}. \cite{parker1996inferring} suggested that these isolated flux tubes are individually current-neutralized, which requires the main currents surrounded by shielding currents of equal amount and in opposite direction. However, \cite{melrose1991neutralized,melrose1995current} argued that net currents can emerge from the solar interior. \cite{longcope2000model} further predicted that most return currents would be trapped below or at the photosphere. 

The answer to current neutralization may have critical consequences for theoretical flare/CME models and pre-eruptive magnetic configuration. While it is a consensus that the key structure of the flare/CME models is a magnetic flux rope (MFR), the nature of the pre-eruptive configuration, a pre-existing MFR or a shear arcade, has been under intense debate. Many previous studies predicted that a current-carrying MFR can emerge from the solar convection zone into the corona, injecting the non-neutralized currents into the ARs (e.g. \citealt{leka1996evidence,titov1999basic,fan2004numerical,fan2009emergence,demoulin2010criteria,cheung2014flux}). \cite{torok2014distribution} modeled the emergence of a sub-photospheric current-neutralized MFR into the solar atmosphere. In their three-dimensional magnetohydrodynamic (MHD) simulation, a strong deviation from current neutralization was found at the end of emergence. Further, \cite{dalmasse2015origin} investigated the distribution and neutralization of currents generated by photospheric horizontal flows in 3D, zero-$\beta$, MHD simulations. In their experiment, net currents would develop around the polarity inversion line (PIL), when photospheric plasma flows produced magnetic shear along the PIL.

Recently, more attention has been paid to investigate these theoretical considerations using uninterrupted high spatial-resolution data on vector magnetic fields provided by space instruments (e.g. the Helioseismic and Magnetic Imager on board Solar Dynamics Observatory, SDO/HMI, \citealt{pesnell2011solar}). \cite{georgoulis2012non} performed a detailed observational study of electric current patterns in two ARs. They found that only the ARs with well-formed PILs contain stronger non-neutralized current patterns per polarity. \cite{liu2017electric} indicated that the degree of current neutralization would be a better proxy for assessing the CME productivity of the ARs by conducting a pilot observational study of four ARs. Later, this relationship has been explored in larger samples \citep{vemareddy2019degree,kontogiannis2019which,avallone2020electric}. \cite{avallone2020electric} had investigated the degree of current neutralization in 30 ARs (15 flare-active and 15 flare-quiet). Their results confirmed that most flare-active ARs own non-neutralized currents, while flare-quiet ARs exhibit the characteristic of neutralization.

Those observations, however, did not clarify whether non-neutralized currents in active regions are necessarily signatures of a magnetic flux rope (MFR) that are present before the eruption. To understand formation mechanisms of non-neutralized currents and verify previous models and simulations, it is necessary to investigate electric currents of MFRs in the observations. However, the technology of direct measuring the coronal magnetic field is still immature, thus it poses a major challenge to observational investigation of MFRs' magnetic properties. According to 3D extension of the standard flare models and numerical models (e.g. \citealt{gosling1990coronal,moore2001onset,janvier2014electric,aulanier2019drifting}), magnetic reconnection occurring around and below the erupting MFR should produce J/Z-shaped flare ribbons, with the ribbon hooks marking the boundary of the photospheric feet of the erupting MFR. \cite{barczynski2020electric} attempted to search possible regions of MFRs' feet based on these characteristics. Unfortunately, less than half flares in their sample were observed with clear ribbon hooks, which are considered as boundary of the footpoints. 
 
To specifically map the MFR's feet, more observational features are required. The ejection of emitting plasma will cause transient darkening of the areas in the eruptive region, named as coronal dimmings or transient coronal hole \citep{harrison2000spectroscopic,harra2001material}. Particularly,  conjugate coronal dimmings that occur in the vicinity of flare ribbons and are located in photospheric fields of opposite polarities can well map the MFR's feet \citep{webb2000relationship,qiu2007magnetic,hu2014structures,cheng2016nature,qiu2017gradual,wang2017buildup,wang2019evolution}. In addition, \cite{qiu2017gradual} and \cite{wang2019evolution} observed two eruptive events with clear conjugate dimmings that appeared several hours before the eruptions. Therefore, conjugate coronal dimmings are a good candidate to help identify MFRs' foot-prints. In particular, if pre-eruptive dimmings are present at the same locations, they will help diagnose the dynamic and magnetic evolution of coronal structures, which are likely MFRs, before the MFR eruption.

Here we conduct a statistical study to explore pre-eruptive magnetic properties of erupting MFRs. We quantify magnetic properties at the feet of 28 MFRs prior to their eruptions using high-quality vector magnetograms from HMI. In the following section, we will briefly introduce the selection of eruptive events, our method of footpoints identification, measurements of magnetic properties at the footprints of MFRs, and uncertainties of MFRs properties estimated in this study. The statistical results will be shown in the Section~\ref{sec:ana} and Section~\ref{sec:non}. Summaries and conclusions are given in Section~\ref{sec:dis}.

\section{Data and Methodology} \label{sec:obs}
\subsection{Dataset}\label{sec:dat}

According to previous flare/CME models (e.g. \citealt{moore2001onset,janvier2014electric}) and observational studies (e.g. \citealt{hu2014structures,cheng2016nature,qiu2017gradual,wang2017buildup,wang2019evolution}), two-ribbon flares with conjugate dimmings will be good candidates for identifying the MFRs' footpoints, and subsequently measuring their magnetic properties. We have examined 400 two-ribbon flares of GOES class C5.0 and larger from a database \textsl{RibbonDB} \citep{kazachenko2017database}, which includes all flares of GOES class C1.0 and larger, observed by the Atomspheric Imaging Assemly (AIA; \citealt{lemen2012atmospheric}) onboard SDO from 2010 to 2016. To minimize projection effects, we choose flare events that occur within 45$^{\circ}$ from the central meridian. Meanwhile, only flares of GOES class larger than C5.0 and smaller than X2.0 are considered. Then only 52 flare events are retained. Our preliminary analysis of these events shows that about 28 events exhibit evident conjugate dimmings. We hence select the 28 events with conjugate dimmings to conduct the statistical study.

Table~\ref{tab1} provides observational properties of the 28 eruptive events. All studied events are associated with CMEs observed by the Solar Terrestrial Relations Observatory (STEREO; \citealt{kaiser2008stereo}) and/or the Large Angle and Spectrometric Coronagraph Experiment (LASCO). From STEREO and LASCO observations, most CMEs are showing classical three-part structures or twisted loop-like structures, implying the existence of MFRs. Half of them are halo CMEs when viewed from Earth, and 7 among them are associated with magnetic clouds (MCs) observed at 1 AU. Before the eruptions, some plasma proxies for MFRs, including sigmoids, hot channels, filaments, and expanding coronal loops, are well visible for most events.

\subsection{Identification of erupting MFRs' footpoints}\label{sec:fp}

Many previous studies \citep{webb2000relationship,qiu2007magnetic,cheng2016nature,qiu2017gradual,wang2017buildup,wang2019evolution} indicated that conjugate dimmings located in magnetic fields of opposite polarities map the MFR's feet. 
Statistical studies of coronal dimmings \citep{dissauer2018detection,dissauer2018statistics,dissauer2019statistics} suggested that coronal dimmings often occur in multiple extended areas, due to projection or interation of the large scale structure of the erupting MFR. Their studies also found that only small parts of the coronal dimming map the feet of the original MFR, and their edges are difficult to identify automatically. Therefore, in the past, combined manual and semi-automatic detection methods were employed to determine feet-related dimming regions 
(e.g. \citealt{qiu2017gradual,wang2017buildup,wang2019evolution,xing2020evolution}).   

In this study, we have developed a novel automated algorithm to detect footpoints of erupting MFRs, based on theoretical concepts and observational characteristics of MFRs. This method has been improved upon our previous 
detection method of coronal dimmings (see \citealt{wang2017buildup,wang2019evolution}). First of all, our method will analyze seven EUV passbands of AIA observations synchronously to seek possible coronal dimming regions. This step will guarantee the detected dimmings are due to mass evacuation related to eruption or expansion, rather than the change in the plasma temperature. Then all detected regions will be projected onto the preflare HMI vector magnetogram to check their magnetic polarity and connectivity. In this process, the dimmed regions that are co-spatial with the magnetic field of mixed polarity and conflicting connectivity are discarded. According to 3D extension of flare/CME models (e.g.\citealt{moore2001onset,janvier2014electric}), feet-related dimmings tend to appear in the vicinity of flare ribbons. In the next step, our method will examine the location of detected dimmings, selecting the dimmed regions that occur along the two flare ribbons. The flare ribbons are detected in the AIA 1600 channel using the method from \cite{qiu2007magnetic}. Finally, two boundaries of conjugate dimmings will be determined. More detailed information can be found in the Appendix~\ref{app:A}.

\subsubsection{Pre-eruption and post-eruption dimmings}

\begin{figure}
	\epsscale{.9}
	\plotone{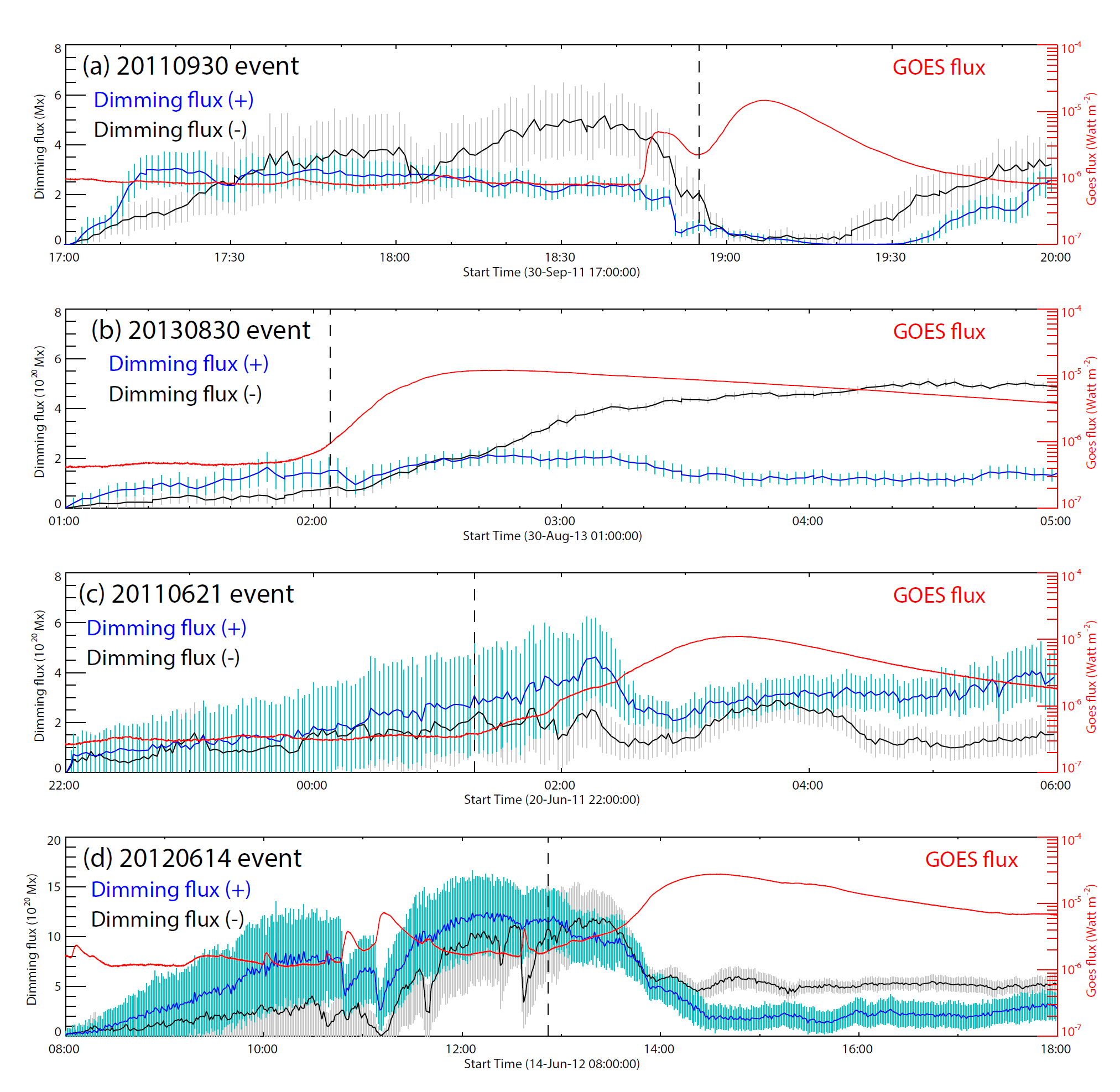}
	\caption{Evolution of pre-eruption coronal dimmings for four events. All diagrams show temporal evolution of dimming fluxes (black for negative, blue for positive) and GOES flux (red). The vertical dashed line represents the onset of the flare for each event. \label{fig1}}
\end{figure}

To estimate the specific areas of the expanding or erupting MFRs' feet, we quantify the evolution of conjugate dimmings. Many studies had indicated that some coronal dimming signatures may be caused by projection effects, when coronal loops evolve and change their orientations  \citep{harvey2002polar,harra2007coronal,qiu2007magnetic,scholl2008automatic}. To minimize such projection effects, we track evolution of dimmings in the AIA 304 passpand. We then project the dimming pixels onto the HMI vector magnetogram half an hour before each flare and sum up the magnetic flux within the dimming regions to calculate the dimming fluxes. The results show that conjugate dimmings will appear before or after the onset of flares, which are termed pre-eruption dimmings (Figure~\ref{fig1}) or post-eruption dimmings (Figure~\ref{fig2}). In this study, the pre-eruption dimmings appear at least one hour before the onset of flare. According to previous studies about conjugate dimming fluxes (e.g. \citealt{qiu2007magnetic,hu2014structures,wang2017buildup,wang2019evolution}), the order of magnitude of dimming fluxes are around $10^{20-22}$ Mx. Here the occurrence of dimmings is defined by the moment when the value of dimming fluxes becomes larger than $10^{19}$ Mx.

Our results show that most pre-eruption dimmings will expand in area before the eruption and shrink during the eruption (Figure~\ref{fig1}). It is worth noting that dimming fluxes are calculated using the same pre-flare HMI magnetogram (half an hour before the onset of the flare). For example, the 20110930 event, dimming fluxes rapidly increase and reach a plateau, which lasts about 1.5 hours. The dimming fluxes start to decrease several minutes before the onset of the flare (vertical dashed line in Figure~\ref{fig1} (a)). But for several events, the dimming fluxes grow slowly before the eruption, e.g. the 20130830 event (see Figure~\ref{fig1} (b)). The negative dimming fluxes in the 20130830 event increase fast after the onset of the flare. For the 20110621 event, dimming fluxes start to rise three hours before the flare and continue to rise during the eruption.

In this study, we neglect pixels undergoing transient dimmings, which are mainly detected during the period of rapid changes; but focus on dimmed pixels that persist for a long time, mainly detected during the relatively stable stage in the light-curve of dimming fluxes. As a result, we identify two stationary dimming regions during the whole eruption process. Despite the evolution of coronal dimmings, the two stationary regions are considered as the 'core' feet of the MFR, from which the magnetic properties of the MFR are investigated. For examples, for the 20110930 and 20120614 event, we select all dimming pixels detected during the plateau of fluxes. For the 20110621 event and 20130830 event, we select all dimming pixels when the fluxes grow to 40\% of it maximum. The identified footpoints of these four events are shown in the Figure~\ref{fig3} (white contours). In our sample of pre-eruption dimmings, we find that, when flare occurs, flare ribbons tend to cover part of dimming areas (see Figure~\ref{fig3} (red contours)), explaining the reduction of dimming fluxes during the eruption.

Our results show that the post-eruption dimmings always experience a stage of rapid expansion in area, followed by a stage of relatively stable and slow growth (Figure~\ref{fig2}). For example, in the 20120310 event, the negative dimming flux increases rapidly after the onset of the flare (Figure~\ref{fig2} (d)). For the post-eruption dimmings, we select all dimming pixels detected after the peroid of rapid growth. Then we consider that these pixels map the MFR footpoints. The white contours in Figure~\ref{fig4} denote the conjugate footpoints of the MFR identified with this method. For post-eruption dimmings, the flare ribbons also cover part of the dimming areas (Figure~\ref{fig4}). In this way, footpoints of 28 MFRs are all well identified.

\begin{deluxetable*}{CC|CC|CC|CC|C|C|C}[!b]
	\tabletypesize{\scriptsize}
	\tablecaption{Overview of Eruptions\label{tab1}}
	\tablecolumns{14}
	\tablenum{1}
	\tablewidth{0pt}
	\tablehead{
		\colhead{} & \colhead{} & \colhead{} & \colhead{} & \colhead{} & \colhead{} & \colhead{} & \colhead{} & \colhead{} & \colhead{} & \colhead{}\\
		\colhead{No.} & \colhead{Date} & \colhead{AR} & \colhead{} & \colhead{Flare} &  \colhead{} & \colhead{Dimming} & \colhead{} & \colhead{CME} & \colhead{MC} & \colhead{MFR identity} \\
		\colhead{} & \colhead{} & \colhead{NOAA} & \colhead{Location} & \colhead{class} & \colhead{Onset}  & \colhead{Onset} & \colhead{Label} & \colhead{} & \colhead{} & \colhead{}
	}
	\startdata
	1 & 20100807 & 11093 & N12E31  & M1.0 & 17:55 &18:35 & Post & Halo & - & Double-Decker \ filament \\
	\hline
	2 & 20110307 & 11166 &  N11E13 & M1.7 & 13:45 & 14:10 & Post & Yes & - & Hot \ channel \\
	\hline
	3 & 20110621 & 11236 & N17W21 & C7.7 & 01:18 & *22:02 & Pre & Yes & - & Sigmoid-like \ filament \\
	\hline
	4 & 20110802 & 11261 & N16W22 & M1.4 & 05:19 &  05:07 & Post & Halo & - & Hot \ channel \\
	\hline
	5 & 20110930 & 11305 & N13E02 & M1.3 & 18:55 & 17:03 & Pre & Yes & - & Expanding \ coronal \ structure\\  
	\hline
	6 & 20120309 & 11429 & N17W13 & M6.3 & 03:22 & 03:40 & Post & Halo & - & Hot \ channel \\
	\hline
	7 & 20120310 & 11429 & N18W27 & M8.4 & 17:15 & 17:30 & Post & Halo & - & Hot \ channel \\
	\hline
	8 & 20120614 & 11504 & S17W00 & M1.0 & 12:52 & 08:00 & Pre & Halo & Yes & Expanding \ coronal \ structure \\
	\hline
	9 & 20120712 & 11520 & S17W08 & X1.4& 15:37 & 16:18 & Post & Halo & Yes & Hot \ channel \\
	\hline
	10& 20130206 & 11667 & N22W00  & C8.7 & 00:04 &  00:10 & Post & Yes & - & Filament  \\
	\hline
	11& 20130411 & 11719 & N10W01 & M6.5 & 06:55 & 06:01 & Pre & Halo & Yes & Hot \ channel \\
	\hline
	12& 20130517 & 11748 & N12E22  & M3.2 & 08:43 & 08:52 & Post & Halo & - & Hot \ channel \\
	\hline
	13& 20130812 & 11817 & S22E10 & M1.5 & 10:21 & 10:42 & Post & Yes & - & \\
	\hline
	14& 20130817 & 11818 & S07W32 & M3.3 & 18:16 &  18:40 & Post & Yes & - & Twisted \ loops \\
	\hline
	15& 20130830 & 11836 & N12E28 & C8.3 & 02:04 &  01:04 & Pre & Yes & - & Expanding \ coronal \ structure \\
	\hline
	16& 20131013 & 11865 & S22E05 & M1.7 & 00:12 & 00:32 & Post & Yes & - & \\
	\hline
	17 & 20140131 & 11968 & N09E29 & M1.1 & 15:32 & 15:40 & Post & Yes & - & \\
	\hline
	18& 20140320 & 12010 & S15E27 & M1.7 & 03:42 & 03:52 & Post & Halo & - &  \\
	\hline
	19& 20140730 & 12127 & S08E34 & C9.0 & 16:00 & 16:10 & Post & Yes & - & \\
	\hline
	20& 20140801 & 12127 & S09E08 & M1.5 & 17:55 & 18:22 & Post & Halo & - & \\
	\hline
	21& 20140825 & 12146 & N09W47 & M2.0 & 14:46 & 14:01 & Pre & Yes & - & Expanding \ coronal \ structure \\
	\hline
	22& 20140825 & 12146 & N09W47 & M3.9 & 20:06 & 19:35 & Pre & Halo & - & Expanding \ coronal \ structure \\
	\hline
	23& 20140908 & 12158 & N16E26 & M4.5& 23:12 & 22:31 & Pre & Halo & Yes & Sigmoid\\
	\hline
	24& 20140910 & 12158 & N15E02 & X1.6 & 17:21 & 16:02 & Pre & Halo & Yes & Sigmoid \\
	\hline
	25& 20140921 & 12166 & N11W55 & C5.2 & 11:31 &11:24 &  Post & Yes & Yes &  \\
	\hline
	26& 20141220 & 12242 & S18W42 & X1.8 & 0:11 & 00:30 & Post & Yes & - & \\
	\hline
	27& 20150622 & 12371 & N13W14 & M6.5 & 17:39 & 17:59 & Post & Halo & - & Hot \ channel\\
	\hline
	28& 20151104 & 12443 & N06W10 & M3.7 & 13:31 & 13:45 & Post & Halo & Yes & \\
	\enddata
	\tablecomments{ Table~\ref{tab1} shows detailed information of 28 flare events. The dimmings are marked as "pre" when it appeared at least half hour before the start time of the associated flare. The dimmings are marked as "post" when it appeared after the onset of flares. * the onset of dimmings occurred one day before the date of the flare.}
\end{deluxetable*}

\clearpage

\subsection{Measuring pre-eruptive magnetic properties of MFRs}

\subsubsection{Magnetic flux}
By projecting the identified footpoints onto the pre-flare HMI vector magnetogram, we can estimate pre-eruptive magnetic properties of erupting MFRs. In this study, we use the Space-weather HMI Active Region Patch (SHARP) data series, hmi.sharp\_cea\_720s \footnote{\url{http://jsoc.stanford.edu/HMI/Vector_products.html}}, which is disambiguated and deprojected to the heliographic coordinates with a Lambert (cylindrical equal area) projection method, resulting in a pixel scale of 0.36 Mm \citep{bobra2014helioseismic}. For each event, we select one pre-flare HMI vector magnetogram, about half an hour before the onset of the flare, to calculate magnetic fluxes and electric currents. The net magnetic flux ($\Phi_{net}$) can be estimated by summing up $B_{z}$ within the identified footpoint regions.  The uncertainties of magnetic fluxes are estimated by error propagations:
	\[\delta_{\Phi} = \sqrt{\sum_{S}^{} (\delta_{B_{z}})^2}\]
where $\delta_{B_{z}}$ is directly taken from the uncertainties of the HMI data \citep{hoeksema2014helioseismic}.

\subsubsection{Electric current}\label{sec:cur}
In theoretical models, for a coherent MFR, the electric current density $\vect{j}$ near its center will flow in one direction, termed as "direct current" (DC), while the $\vect{j}$ around the MFR periphery must flow in the opposite direction, termed as "return current" (RC) \citep{liu2017electric,sun2020non}. The vertical current density $j_{z}$ and the net current $I_{z}$ can be estimated from the HMI vector magnetogram using the Ampere's law and Stokes theorem:
\[ j_{z}=(\bigtriangledown \times \vect{B})_{z} /  \mu_{0} \]
\[I_{z}=\frac{1}{\mu_{0}}\int_{S}j_{z}dS = \frac{1}{\mu_{0}}\oint_{C}B_{h}\cdot dl\]
where $S$ is the area of identified footpoints, $C$ is the perimeter of $S$, and $\mu_{0}=4\pi \times 10^{-7} H m^{-1}$. 

Following the calculation of current neutralization from \cite{liu2017electric,avallone2020electric,sun2020non}, the index $R_{z} = |I_{z}^{DC}/I_{z}^{RC}|$ is utilized to determine whether the MFR contains non-neutralized currents or not. For each event, $I_{z}^{DC}$ and $I_{z}^{RC}$ are computed for the MFR's footpoints by integrating $j_{z}$ values of a different sign separately. In order to determine the sign of $I_{z}^{DC}$, we search for dominate sign of $j_{z}/B_{z}$ in the MFR's feet. Figure~\ref{fig5} shows electric current density maps for four events as examples. For the 20110930 and 20120614 events, the sign of $I_{z}^{DC}$ for each foot is very obvious. But for the 20120310 and 20140921 event, it will be difficult to determine the sign of $I_{z}^{DC}$.

The HMI vector field has an unavoidable 180$^{\circ}$ ambiguity in the transverse field direction. For the disambiguation of HMI data, the minimum energy algorithm \citep{metcalf1994resolving,metcalf2006overview,leka2009resolving}, which is based on a linear force-free field, is employed. This method may not well disambiguate for weak field regions, where the signal is dominated by noise. That will directly affect our calculation of current neutralization, since many footpoints locate in the weak-field regions. In some previous studies (e.g. \citealt{avallone2020electric}), they only considered the regions where the absolute value of the magnetic field is stronger than 200 G. But the value of electric current will be dramatically reduced when removing weak-field pixels. More detailed comparison between original electric currents and the currents that are calculated by only sum up strong-field pixels can be found in the Appendix~\ref{app:B}.

After many attempts, we find that averaging $j_{z}$ over several hours will largely minimize the effect of low signal-to-noise pixels in the calculation. Comparing with the original electric current density maps (Figure~\ref{fig5} (a2) to (d2)), most weak-field pixels are eliminated in the average maps (Figure~\ref{fig5} (a3) to (d3)). To further display temporal evolution of these pixels, we cut out four weak-field regions (50$\times$50 pixel) from four events (see four squares in the Figure~\ref{fig5}). For the weak-field pixels, the plus or minus sign of $j_{z}$ always appear alternately. Therefore, averaging current density maps over a period of time will well eliminate these noise-like pixels. In this study, we average $j_{z}$ over a period of two hours, noting that the timescale of the evolution of the photospheric field is several hours.

The results indicate two MFR populations, one carrying net currents (at least one foot with the value of $R_{z}$ larger than 2.0), and the other carrying neutralized currents. We find 8 out of 28 MFRs contain significant non-neutralized currents. For example, the $R_{z}$ for one foot of 20110930 event is about 7.0, while the $R_{z}$ in the two feet of the 20140921 event ranges from 1.0 to 1.2. The histogram of $R_{z}$ for the 28 MFRs is shown in the Figure~\ref{fig6}. For each foot, the number of events first decreases with the increasing $R_{z}$, and then flattens at around 1.8 to 2.0. Previous studies provided the $R_{z}$ for the active region ranging from 1.1 to 2.0 \citep{liu2017electric,avallone2020electric}. We suggest $1.8-2.0$ as an possible empirical threshold to distinct the MFRs with or without net currents when considering the average value of $R_{z}$ in two footpoints.

Similar to magnetic fluxes, the uncertainties of electric currents are also estimated by error propagations:
\[\delta_{I} = \sqrt{\sum_{S}^{} (\delta_{j_{z}})^2}\]
But electric currents are calculated from the aforementioned average $j_{z}$ maps. Then we consider the standard deviation of the variation of $j_{z}$ during two hours as $\delta_{j_{z}}$. More information about the uncertainties of $j_{z}$ can be found in the Appendix~\ref{app:B}. For the $\delta_{I_{z}^{DC}}$ and $\delta_{I_{z}^{RC}}$, we only consider pixels carrying the current with the same polarity of $I_{z}^{DC}$ or $I_{z}^{RC}$. The uncertainties of the degree of current neutralization ($\delta_{R_{z}}$) is also given by error propagations:
\[ \delta_{R_{z}} = R_{z} \sqrt{ (\frac{\delta_{I_{z}^{DC}}}{I_{z}^{DC}})^2 + (\frac{\delta_{I_{z}^{RC}}}{I_{z}^{RC}})^2 }\]. 

\clearpage

\section{The pre-eruptive magnetic properties of erupting MFRs}\label{sec:ana}

Table~\ref{tab2} and Table~\ref{tab3} list the properties measured at two feet of each of the 28 MFRs, including the net magnetic flux ($\Phi_{net}$), the net electric current ($I_{z}^{net}$), the direct current ($I_{z}^{DC}$), the return current ($I_{z}^{RC}$), and the degree of current neutralization ($R_{z}$). For each event, magnetic fluxes are estimated from the HMI vector magnetogram obtained half an hour before the eruption, while electric currents are calculated using the current density map averaging over two hours before the eruption. The $\Phi_{net}$ of MFRs is around $10^{20-21}$ Mx, comparable to toroidal fluxes calculated from magnetic clouds \citep{hu2014structures}. The $I_{z}^{net}$ of MFRs is around $10^{11-12}$ A, in the same order of magnitude as the current calculated from ends of sigmoids \citep{cheng2016characteristics}. The $I_{z}^{DC}$ in their study is about an order of magnitude higher than ours, probably because they arbitrarily choose a rectangular region as the footpoints. In addition, the $I_{z}$ in our study is derived by averaging the measurements in two hours. In the tables, the sign of footpoints represents the sign of magnetic polarities, positive (+) or negative (-). Furthermore, we also distinguish the two feet as the leading footpoint (L) and the trailing footpoint (T), according to the direction of solar rotation.

\subsection{The footpoints identified by pre-eruption conjugate dimmings}
In our sample of 28 events, we find 9 events with obvious pre-eruption conjugate dimmings, which appear at least one hour before the onset of the flare (see Table~\ref{tab1}). The observational signatures of the MFR, e.g. expanding coronal structures \citep{wang2019evolution}, sigmoid-like filaments \citep{zhou2017toward}, sigmoids and hot channels \citep{cheng2016characteristics}, are observed for each event before the eruption. For these events, conjugate dimmings evolve simultaneously with the MFR-like structures, suggesting that the pre-eruption dimmings map the feet of the MFR-like structures.

Here we show a typical case about the pre-eruption dimmings, the 20120614 event. The eruption of interest is associated with an M1.9 flare, which is captured by SDO and STEREO. The observations from STEREO-B/EUVI reveal a gradual expansion of a coronal structure lasting for more than five hours before the onset of the flare (Figure~\ref{fig7} (a1) to (a3)), which finally evolved into a halo CME (Figure~\ref{fig7} (b)). The average brightnesses for the two identified footpoint regions evolve simultaneously with the expanding coronal structure (Figure~\ref{fig7} (c) and (d)). More detailed investigation about this event can be found in our previous study \citep{wang2019evolution}. In our sample, most pre-eruption dimmings evolve simultaneously with expanding structures, which finally erupt as a CME.

We further compare dimming areas detected before and after the onset of flares. For most events, dimming areas detected before the flare are always larger than dimming areas detected after the flare. Figure~\ref{fig8} shows the comparison for the aforementioned four events. For 20110930 and 20120614 event, dimming areas detected before the flare (blue contours in the Figure~\ref{fig8} (a) and (d)) completely cover dimming areas detected after the flare (green contours in the Figure~\ref{fig8} (a) and (d)). For 20110621 event, two areas are nearly the same (Figure~\ref{fig8} (c)). For 20130830 event, as one exception, the dimming area after the flare is larger than the area before the flare. More interestingly, for 20110930 event, flare ribbons occur inside two dimming areas (Figure~\ref{fig3} (a)), causing the reduction of dimming fluxes (Figure~\ref{fig1} (a)). Yet after the flare, dimming occurs again in the region covered by flare ribbons, hence the dimming fluxes grow again after the flare. (Figure~\ref{fig1} (a)).

Table~\ref{tab2} shows quantitative measurements of magnetic properties within the footpoints identified by pre-eruption conjugate dimmmings. The $|\Phi_{net}|$ in the footpoints range from $3.62\times10^{20}$ to $50.95\times10^{20}$ Mx and the  $|I_{z}^{net}|$ range from $0.14\times10^{11}$ to $19.12\times10^{11}$ A. The maximum of $|I_{z}^{DC}|$ is about $41.91\times10^{11}$ A, and the minimum is about $2.65\times10^{11}$ A, much higher than the range of $|I_{z}^{net}|$. But the range of $R_{z}$ is from 1.0 to 7.0. The results show that four events are associated with non-neutralized currents ($R_{z}>2.0$), marked in bold font in Table~\ref{tab2}. The $R_{z}$ of the other events are around 1.0 to 1.5.

\begin{deluxetable*}{C|CC|CCCCCC}[b!]
	\tabletypesize{\scriptsize}
	\tablecaption{The pre-eruptive magnetic properties of pre-eruption dimming events \label{tab2}}
	\tablecolumns{10}
	\tablenum{2}
	\tablewidth{0pt}
	\tablehead{ 
		\colhead{No.} & \colhead{Date} & \colhead{FP} & \colhead{} & \colhead{$\Phi_{net}$} & \colhead{$I_{z}^{net}$} & \colhead{$I_{z}^{DC}$} & \colhead{$I_{z}^{RC}$} & \colhead{$R_{z}$} \\
		\colhead{} & \colhead{} & \colhead{Sign} & \colhead{Type} &  \colhead{($10^{20}$ Mx)} & \colhead{($10^{11}$ A)} & \colhead{($10^{11}$ A)} & \colhead{($10^{11}$ A)} & \colhead{}
	}
	\startdata
	3 & 20110621 & + & T & 13.56$\pm$0.02 & 0.53$\pm$1.82 & 30.17$\pm$1.26 & -29.64$\pm$1.24 &  1.0$\pm$0.1\\
	&          & - & L & -12.70$\pm$0.03 & -1.60$\pm$1.68 & -23.43$\pm$1.07 & 21.83$\pm$1.06 & 1.1$\pm$0.1 \\
	\hline
	\bm{5} & \bm{20110930} & + & T & 5.42$\pm$0.01 & -4.38$\pm$1.49 & -16.33$\pm$0.91 & 11.95$\pm$0.78 & 1.4$\pm$0.1 \\
	&          & \bm{-} & \bm{L} & -13.77$\pm$0.21 & 7.64$\pm$1.06 & 8.92$\pm$0.42 & -1.28$\pm$0.21 & \bm{7.0$\pm$1.2} \\
	\hline
	\bm{8} & \bm{20120614} & \bm{+} & \bm{L} & 40.00$\pm$0.03 & 11.52$\pm$0.87 & 13.60$\pm$0.28 & -2.08$\pm$0.15 & \bm{6.5$\pm$0.5}\\
	&          & \bm{-} & \bm{T} & -30.67$\pm$0.02 & -19.12$\pm$1.32 & -31.47$\pm$0.70 & 12.35$\pm$0.61 & \bm{2.6$\pm$0.1}\\
	\hline
	11& 20130411 & + & T & 8.62$\pm$0.03 & -0.62$\pm$1.80 & -2.65$\pm$1.23 & 2.58$\pm$1.21 & 1.0$\pm$0.7 \\
	&          & - & L & -5.91$\pm$0.01 & 0.79$\pm$0.93 & 5.84$\pm$0.32 & -5.05$\pm$0.36 & 1.2$\pm$0.1\\
	\hline
	15& 20130830 & + & T & 21.24$\pm$0.08 & 2.51$\pm$1.61 & 23.04$\pm$0.99 & -20.53$\pm$0.96 & 1.1$\pm$0.1\\
	&          & - & L & -15.85$\pm$0.07 & 0.14$\pm$2.08 & 44.71$\pm$1.64 & -44.57$\pm$1.64 & 1.0$\pm$0.1\\
	\hline
	\bm{21}& \bm{20140825} & \bm{+} & \bm{T} & 4.21$\pm$0.02 & -4.54$\pm$1.04 & -8.45$\pm$0.45 & 3.91$\pm$0.37 & \bm{2.2$\pm$0.2}\\
	&          & \bm{-} & \bm{L} & -6.24$\pm$0.01 & 3.53$\pm$0.82 & 6.74$\pm$0.26 & -3.21$\pm$0.26 & \bm{2.1$\pm$0.2} \\
	\hline
	22& 20140825 & + & T & 3.62$\pm$0.02 & -0.84$\pm$1.19 & -7.41$\pm$0.55 & 6.57$\pm$0.52 & 1.1$\pm$0.1\\
	&          & - & L & -5.25$\pm$0.01 & 1.18$\pm$0.84 & 4.62$\pm$0.27 & -3.44$\pm$0.28 & 1.3$\pm$0.1\\
	\hline
	\bm{23}& \bm{20140908} & \bm{+} & \bm{L} & 50.95$\pm$0.04 & -16.75$\pm$1.01 & -23.71$\pm$0.44 & 6.96$\pm$0.38 & \bm{3.4$\pm$0.2}\\
	&          & - & T & -44.39$\pm$0.04 & 8.01$\pm$1.69 & 41.91$\pm$1.13 & -33.93$\pm$1.02 & 1.2$\pm$0.1\\
	\hline
	24& 20140910 & + & L & 19.64$\pm$0.02 & -5.71$\pm$1.23 & -16.47$\pm$0.50 & 10.76$\pm$0.57 & 1.5$\pm$0.1\\
	&          & - & T & -29.38$\pm$0.02 & 10.50$\pm$1.55 & 33.48$\pm$0.95 & -22.99$\pm$0.87 & 1.5$\pm$0.1\\
	\hline
	\enddata
	\tablecomments{Table~\ref{tab2} shows the properties of pre-eruption dimming events. The '+'/'-' in the table represents the footpoints of postive/negative magnetic polarities respectively, and 'L/T' represent the leading/trailing footpoints. The footpoints with non-neutralized current are in bold font.}
\end{deluxetable*}

\subsection{The footpoints identified by post-eruption conjugate dimmings}
The rest 19 events are all associated with obvious post-eruption conjugate dimmings (see Table~\ref{tab1}). For half of the events, we did not find obvious signatures of MFRs before the eruption. For most events, dimmings appear several minutes or dozens of minutes after the onset of flares. The dimmings always undergo a rapid rise phase followed by a long stable phase (see Figure~\ref{fig2}). As flare/CME models predicted, most post-eruption conjugate dimmings appear in the two end of flare ribbons. For example, the 20140921 event, two dimmings are located along the two flare ribbons (Figure~\ref{fig4} (b)). Some dimmings even appear inside the flare ribbons. For example, the 20120310 event, flare ribbons cover the most part of dimming areas (Figure~\ref{fig4} (d)).

Table~\ref{tab3} shows quantitative measurements of magnetic properties within the footpoints identified by post-eruption conjugate dimmmings. The $|\Phi_{net}|$ in the footpoints range from $1.96\times10^{20}$ to $42.74\times10^{20}$ Mx and the  $|I_{z}^{net}|$ range from $0.02\times10^{11}$ to $11.52\times10^{11}$ A. The maximum of $|I_{z}^{DC}|$ is about $55.49\times10^{11}$ A, and the minimum is about $3.42\times10^{11}$ A, much higher than the range of $|I_{z}^{net}|$. These values  are similar to the results from the pre-eruption dimming events. But the range of $R_{z}$ is smaller than that of the pre-eruption dimming events, from 1.0 to 3.6. Only 4 of 19 events are associated with non-neutralized currents ($R_{z}>2.0$), marked in bold font in Table~\ref{tab3}. The $R_{z}$ for most events are around 1.0 to 1.6.

\begin{deluxetable*}{C|C|CC|CCCCC}[b!]
	\tabletypesize{\scriptsize}
	\tablecaption{The pre-eruptive magnetic properties of post-eruption dimming events\label{tab3}}
	\tablecolumns{10}
	\tablenum{3}
	\tablewidth{0pt}
	\tablehead{ 
		\colhead{No.} & \colhead{Date} & \colhead{FP} & \colhead{} & \colhead{$\Phi_{net}$} & \colhead{$I_{z}^{net}$} & \colhead{$I_{z}^{DC}$} & \colhead{$I_{z}^{RC}$} & \colhead{$R_{z}$} \\
		\colhead{} & \colhead{} & \colhead{Sign} & \colhead{Type} & \colhead{($10^{20}$ Mx)} & \colhead{($10^{11}$ A)} & \colhead{($10^{11}$ A)} & \colhead{($10^{11}$ A)} & \colhead{}
	}
	\startdata
	1 & 20100807 & + & T & 4.90$\pm$0.02 & -0.11$\pm$1.31 & -8.67$\pm$0.64 & 8.56$\pm$0.65 & 1.0$\pm$0.1\\
	&          & - & L & -6.55$\pm$0.02 & -0.02$\pm$1.15 & -7.32$\pm$0.54 & 7.30$\pm$0.50 & 1.0$\pm$0.1\\
	\hline
	2 & 20110307 & + & T & 4.62$\pm$0.02 & -2.86$\pm$1.31 & -10.31$\pm$0.65 & 7.45$\pm$0.64 & 1.4$\pm$0.2 \\
	&          & - & L & -7.01$\pm$0.02 & 0.21$\pm$1.29 & 10.18$\pm$0.62 & -9.97$\pm$0.60 & 1.0$\pm$0.1\\
	\hline
	\bm{4} & \bm{20110802} & \bm{+} & T & 13.62$\pm$0.01 & 11.52$\pm$0.91 & 15.93$\pm$0.36 & -4.41$\pm$0.27 & \bm{3.6$\pm$0.2}\\
	&          & - & L & -8.00$\pm$0.02 & -4.28$\pm$1.16 & -11.68$\pm$0.64 & 7.39$\pm$0.50 & 1.6$\pm$0.1\\
	\hline
	6 & 20120309 & + & L & 17.23$\pm$0.03 & -5.34$\pm$1.48 & -19.77$\pm$0.82 & 14.43$\pm$0.80 & 1.4$\pm$0.1\\
	&          & - & T & -17.58$\pm$0.03 & 7.35$\pm$1.45 & 21.69$\pm$0.87 & -14.34$\pm$0.79 & 1.5$\pm$0.1\\
	\hline
	7 & 20120310 & + & L & 10.31$\pm$0.04 & 2.19$\pm$1.61 & 23.84$\pm$0.97 & -21.65$\pm$0.91 & 1.1$\pm$0.1\\
	&          & - & T & -30.79$\pm$0.03 & 8.44$\pm$1.46 & 26.87$\pm$0.82 & -18.43$\pm$0.78 & 1.5$\pm$0.1\\
	\hline
	9 & 20120712 & + & L & 42.74$\pm$0.03 & 2.41$\pm$1.87 & 55.49$\pm$1.34 & -53.08$\pm$1.32 & 1.1$\pm$0.0\\
	&          & - & T & -34.46$\pm$0.04 & -0.86$\pm$1.84 & -49.74$\pm$1.26 & 48.88$\pm$1.28 & 1.0$\pm$0.0\\
	\hline
	10& 20130206 & + & T & 14.03$\pm$0.03 & 2.38$\pm$1.41 & 15.36$\pm$0.73 & -12.98$\pm$0.76 & 1.2$\pm$0.1 \\
	&          & - & L & -3.76$\pm$0.02 & -1.81$\pm$1.42 & -12.40$\pm$0.78 & -10.59$\pm$0.76 & 1.2$\pm$0.1\\
	\hline
	12& 20130517 & + & L & 11.90$\pm$0.02 & -1.95$\pm$1.09 & -10.37$\pm$0.45 & 8.42$\pm$0.45 & 1.2$\pm$0.1\\
	&          & - & T & -11.39$\pm$0.05 & 1.42$\pm$1.78 & 30.15$\pm$1.19 & -28.74$\pm$1.19 & 1.1$\pm$0.1\\
	\hline
	\bm{13}& \bm{20130812} & + & L & 3.42$\pm$0.01 & -1.78$\pm$0.79 & -3.95$\pm$0.24 & 2.17$\pm$0.23 & 1.8$\pm$0.2\\
	&          & \bm{-} & \bm{T} & -4.81$\pm$0.01 & 2.43$\pm$0.86 & 3.43$\pm$0.32 & -1.01$\pm$0.24 & \bm{3.4$\pm$0.9}\\
	\hline
	14& 20130817 & + & L & 2.03$\pm$0.01 & 1.58$\pm$0.93 & 4.12$\pm$0.30 & -2.53$\pm$0.33 & 1.6$\pm$0.2\\
	&          & - & T & -8.16$\pm$0.02 & -0.48$\pm$1.19 & -8.62$\pm$0.54 & 8.13$\pm$0.53 & 1.1$\pm$0.1\\
	\hline
	\bm{16}& \bm{20131013} & + & L & 5.82$\pm$0.02 & 2.77$\pm$1.14 & 8.40$\pm$0.49 & -5.63$\pm$0.47 & 1.5$\pm$0.2\\
	&          & \bm{-} & \bm{T} & -8.87$\pm$0.02 & -9.58$\pm$1.23 & -16.39$\pm$0.63 & 6.81$\pm$0.51 & \bm{2.4$\pm$0.2}\\
	\hline
	17& 20140131 & + & T & 2.47$\pm$0.03 & -0.06$\pm$1.44 & -8.47$\pm$0.80 & 8.41$\pm$0.76 & 1.0$\pm$0.1\\
	&          & - & L & -1.96$\pm$0.02 & 0.06$\pm$1.39 & 8.73$\pm$0.73 & -8.67$\pm$0.70 & 1.0$\pm$0.1\\
	\hline
	18& 20140320 & + & L & 8.22$\pm$0.02 & 0.47$\pm$1.27 & 9.96$\pm$0.60 & -9.49$\pm$0.61 & 1.1$\pm$0.1\\
	&          & - & T & -12.90$\pm$0.02 & -1.92$\pm$0.93 & -6.18$\pm$0.35 & 4.26$\pm$0.31 & 1.5$\pm$0.1\\
	\hline
	19& 20140730 & + & L & 10.47$\pm$0.02 & -1.02$\pm$1.18 & -12.04$\pm$0.56 & 11.02$\pm$0.50 & 1.1$\pm$0.1\\
	&          & - & T & -8.19$\pm$0.02 & 0.89$\pm$1.31 & 8.25$\pm$0.65 & -7.36$\pm$0.70 & 1.1$\pm$0.1\\
	\hline
	20& 20140801 & + & L & 8.80$\pm$0.01 & 1.52$\pm$1.30 & 13.56$\pm$0.46 & -12.04$\pm$0.67 & 1.1$\pm$0.1\\
	&          & - & T & -6.72$\pm$0.01 & 1.15$\pm$1.13 & 7.27$\pm$0.50 & -6.12$\pm$0.63 & 1.2$\pm$0.2\\
	\hline
	25& 20140921 & + & T & 4.31$\pm$0.02 & 1.08$\pm$1.19 & 7.67$\pm$0.55 & -6.59$\pm$0.52 & 1.2$\pm$0.1\\
	&          & - & L & -5.64$\pm$0.01 & 0.05$\pm$0.89 & 3.98$\pm$0.30 & -3.93$\pm$0.29 & 1.0$\pm$0.1\\
	\hline
	26& 20141220 & + & L & 15.87$\pm$0.02 & 3.01$\pm$1.55 & 19.08$\pm$0.90 & -16.08$\pm$0.86 & 1.2$\pm$0.1\\
	&          & - & T & -33.19$\pm$0.03 & -2.81$\pm$1.42 & -29.58$\pm$0.77 & 26.77$\pm$0.76 & 1.1$\pm$0.0\\
	\hline
	\bm{27}& \bm{20150622} & \bm{+} & \bm{T} & 19.74$\pm$0.02 & -8.85$\pm$1.04 & -17.63$\pm$0.46 & 8.79$\pm$0.35 & \bm{2.0$\pm$0.1}\\
	&          & - & L & -27.67$\pm$0.02 & 5.77$\pm$1.69 & 32.75$\pm$1.02 & -26.98$\pm$1.08 & 1.2$\pm$0.1\\
	\hline
	28 & 20151104 & + & T & 8.93$\pm$0.01 & 1.50$\pm$1.33 & 12.45$\pm$0.67 & -10.97$\pm$0.66 & 1.1$\pm$0.1\\
	&          & - & L & -8.07$\pm$0.01 & -1.01$\pm$1.88 & -23.57$\pm$1.33 & 22.55$\pm$1.36 & 1.1$\pm$0.1\\
	\enddata
	\tablecomments{Table~\ref{tab3} shows the pre-eruptive magnetic properties of post-eruption dimming events. Similar to the Table~\ref{tab2} }
\end{deluxetable*}

\section{The characteristic of MFR with non-neutralized current}\label{sec:non}

In the entire sample of 28 events, only 8 events are associated with significant non-neutralized currents before eruptions (bold font in Table~\ref{tab2} and Table~\ref{tab3}). Direct comparison between properties of the MFRs with or without net currents are given in the Figure~\ref{fig9} and Figure~\ref{fig10}. For both signs of footpoints, we find a high correlation between the magnetic flux $\Phi_{net}$ in the two footpoints (Figure~\ref{fig9}(a1)), the cross-correlation coefficient being 0.82 for the whole sample, 0.94 for the MFRs with non-neutralized currents, and 0.72 for the MFRs with neutralized currents. When considering the leading/trailing footpoint, the coefficient is slightly higher. The ratio of the fluxes in the two footpoints is close to one. We also find the direct currents $I_{z}^{DC}$ in the two footpoints are comparable and strongly correlated (Figure~\ref{fig9}(a2)). These results suggest that our method has successfully identified conjugate footpoints of MFRs. However, when considering the leading/trailing footpoints, the coefficient for the direct currents becomes smaller, especially for the MFRs with non-neutralized currents (Figure~\ref{fig9}(b2)). Larger deviations are found in the net currents $I_{z}^{net}$ for the MFRs with non-neutralized currents (Figure~\ref{fig9}(a3)(b3)) and the degree of current neutralization $R_{z}$ for all MFRs (Figure~\ref{fig9}(a4)(b4)).

The distributions of magnetic properties from two MFR categories, with or without net current, is shown in the Figure~\ref{fig10}. No statistically significant difference is found between two populations in magnetic fluxes and direct currents. When considering the leading/trailing footpoints, the results are similar. But the distribution of the net current $I_{z}^{net}$ is quite different between two populations. Most $I_{z}^{net}$ for the MFRs with non-neutralized currents (blue part in the Figure~\ref{fig10} (b1) to (b4)) are larger than $5\times10^{11}$ A, while $I_{z}^{net}$ for the MFRs with neutralized currents (red part in the Figure~\ref{fig10} (b1) to (b4)) distributes in $1\times10^{11}$ to $5\times10^{11}$ A. Moreover, similar distributions are found in the leading and trailing footpoints for $I_{z}^{net}$, especially for the MFRs with non-neutralized currents.

In our sample, the footpoints of MFRs carrying non-neutralized currents are co-spatial with strong-field regions ($|B|>500$ G). More interestingly, high electric current densities concentrate on the PIL of the host active regions, manifested as two ribbons of opposite sign. Three typical events are shown in Figure~\ref{fig11}. For these events, the identified feet are anchored at the far end of two current ribbons, as the three-dimensional flare-CME model (e.g. \citealt{janvier2014electric}) predicted. However, the imbalance of $R_{z}$ in two footpoints are found for most MFRs (Figure~\ref{fig9}(a4)(b4)). For example, the 20110802 event, the $R_{z}$ of FP+ is about 3.6, which is more than twice the $R_{z}$ of FP- (about 1.6). In the following subsections, we further investigate asymmetric features inside the MFRs with non-neutralized current.

\subsection{Asymmetric electric current distribution}\label{com}

The larger deviation of $R_{z}$ is associated with asymmetric electric current distributions within two footpoints. Figure~\ref{fig12} shows two representative events, 20110930 and 20120614 events. For the 20110930 event, high current densities concentrate on the part of the sunspot, manifested as a spiral-like ribbon (see Figure~\ref{fig12} (a)). One foot of the MFR (FP-) covers the spiral-like current ribbon. But no obvious feature of electric current is found in the other foot (FP+). The $R_{z}$ of FP- is as high as around 7.0, while the $R_{z}$ of FP+ is only about 1.4. Similarly, a semiarc-like current ribbon is observed at the southeastern boundary of FP- in the 20120614 event (see Figure~\ref{fig12}(b)). However, other footpoint of the MFR (FP+), which is anchored in the sunspot, carries relatively uniform current. The $R_{z}$ of FP- is up to 6.5, while the $R_{z}$ for FP+ is about 2.6. 

More interestingly, for three events (20110930, 20120612 and 20140908), their leading footpoints are all co-spatial with the leading sunspots (see Figure~\ref{fig12}). The value of $R_{z}$ in these leading footpoints are higher then in the trailing footpoints (see bold font in the Table~\ref{tab2}). Moreover, these three events are all associated with pre-eruption dimmings. For example, the 20140908 event, the $R_{z}$ is about 3.4 in the leading footpoint, while the $R_{z}$ is about 1.2 in the trailing one. On the contrary, for other four events (20110802, 20130812, 20131013 and 20150622), which are associated with post-eruption dimmings, the value of $R_{z}$ in their trailing footpoints are higher than in the leading footpoints (see bold font in Table~\ref{tab3}). For example, the 20110802 event, the $R_{z}$ is about 3.6 in the trailing footpoint, while the $R_{z}$ is about 1.6 in the leading one.

But the discrepancy is reduced when considering electric currents ($I_{z}^{net}$ and $I_{z}^{DC}$) inside these MFRs. For most events, the values of $I_{z}^{net}$ and $I_{z}^{DC}$ for one foot are about 2 to 3 times than the other one. Furthermore, for the three events with pre-eruption dimmings (20110930, 20120614 and 20140908), the values of $I_{z}^{DC}$ in the leading footpoints with very high $R_{z}$ (up to 7.0) are smaller than in the trailing footpoints with relatively low $R_{z}$ (around 1.2 to 2.6). For other two events with post-eruption dimmings (20110802 and 20131013), the values of $I_{z}^{DC}$ in the trailing footpoints with higher $R_{z}$ are larger than in the leading footpoints with lower $R_{z}$.

\subsection{Asymmetric magnetic twist}

For the MFRs carrying non-neutralized currents, both magnetic fluxes and direct currents in two footpoints are on the same order of magnitude and of opposite signs, in agreement with the scenario of current-carrying flux rope model. Therefore, we can further investigate their magnetic twist using three simple assumptions. Following our previous study \citep{wang2019evolution}, three different methods based on two assumptions, an axial symmetric cylindrical flux rope and nonlinear force-free magnetic configuration, are employed to estimate the average twist of MFRs. These methods were fully discussed in our previous study \citep{wang2019evolution}. In short, the average twist can be estimated by the following equations:
\[T=\frac{LB_{\theta}(r)}{2\pi r B_{z}(r)} \qquad \qquad (Tw1)\] 
\[T=\int_{0}^L \ \frac{  \mathrm{\mu_{0}}\mathrm{J_{\parallel}}}{ \mathrm{4} \pi \mathrm{B}} \ \mathrm{d}l \qquad \qquad (Tw2)\] 
where L is the length of the MFR axis, r is the distance to the axis. In this study, the geometric center of each foot is considered as the axis. L is estimated from the distance between centers of its feet, by assuming a circular-arc shape of the MFR. For the 8 MFRs with non-neutralized currents, the distance between conjugate feet varies from 26.7 Mm to 97.1 Mm, resulting in the length of MFRs ranging from 42.0 Mm to 152.5 Mm. For these two methods (Tw1 and Tw2), the twist per unit length $\tau$ at each foot is calculated from every pixel in FP+ and FP-. Then the average twist at each foot can be estimated as $\langle \tau \rangle$L. The uncertainties in the $\tau$ measurement, estimated from the error propagation, is very small compared with the standard deviation of $\tau$ at each foot. Therefore, we use the standard deviation of $\tau$ to estimate the uncertainty of the twist ($\delta_{Tw}$) measurement.

Instead of measuring $\tau$ at each pixel and taking the average over all pixels, magnetic fluxes and electric currents in each foot can be directly used to calculate the average twist:
\[ T = \frac{\mathrm{\mu_{0}} \mathrm{I}}{ \mathrm{4} \pi \Phi} \mathrm{L} \qquad \qquad (Tw3)\]
The average twist calculated in this method will be very small when using $I_{z}^{net}$. In comparison, we also use the $I_{z}^{DC}$ in each foot to estimate the average twist. These two different measurements are labeled as $Tw3_{net}$ and $Tw3_{DC}$, respectively. For this method, the uncertainty comes from $\delta_{\Phi}$ and $\delta_{I}$, which are estimated through error propagation.

The measurements from the above-mentioned three methods provide a possible range of twist in the MFRs. The results are shown in Table~\ref{tab4}. For the first method, the average twist of the MFRs is around 1.0 to 3.0 turns before the eruption, close to the critical value for kink instability \citep{hood1979kink,torok2004ideal}. In comparison, the twist calculated by the another two methods is smaller, around 0.5 to 2.5 turns. When considering $I_{z}^{net}$, the twist calculated from the third method will be very small, below 1.0 turn. Despite discrepancies in exact values, all three methods confirm that the average twist of these MFRs is around 1.0 turn before the eruption.

Similar to electric current distributions, the twist within two feet is not symmetric before the eruption. A direct comparison between magnetic twists of two footpoints is given in the Figure~\ref{fig13}. We find low correlations between the average twists calculated in the two feet using the three different methods, the cross-correlation coefficient being 0.22 for Tw1, 0.39 for Tw2, 0.18 for $Tw3_{net}$ and 0.19 for $Tw3_{DC}$. The coefficient will be very low (0.06) for Tw1 when considering the leading/trailing footpoints. The average twist in one foot is about 1 to 2 times than in the other foot, especially for Tw2 and Tw3. The discrepancy becomes larger in $Tw3_{DC}$. 

The first method shows that the average twist in the most leading footpoint is larger than in the trailing footpoint. For the MFRs with post-eruption dimmings, the results from the second and third methods also exhibit that the leading foot is associated with higher twist. But the opposite result is found in the MFRs with pre-eruptions dimmings when considering Tw2 and Tw3.  For example, the 20110930 event, the value of Tw2 in the trailing footpoint is around 1.9, which is more than twice in the leading footpoint.

\begin{deluxetable*}{CC|C|C|C|C|C|CCCC}[b!]
	\tabletypesize{\scriptsize}
	\tablecaption{The average twist in the MFRs with non-neutralized current before eruptions\label{tab4}}
	\tablecolumns{14}
	\tablenum{4}
	\tablewidth{0pt}
	\tablehead{
		\colhead{No.} & \colhead{Date} & \colhead{Distance} & \colhead{Length} & \colhead{Dimming} & \colhead{FP} & \colhead{} & \colhead{Tw1} & \colhead{Tw2} & \colhead{$Tw3_{net}$} & \colhead{$Tw3_{DC}$} \\
		\colhead{} & \colhead{} & \colhead{(Mm)} &  \colhead{(Mm)} & \colhead{label} & \colhead{Sign} & \colhead{Type}  & \colhead{} & \colhead{} & \colhead{} & \colhead{}
	}
	\startdata
	4 & 20110802 & 64.8 & 101.7 & Post & + & T  & 1.8$\pm$0.6 & 1.5$\pm$0.6 & 0.8$\pm$0.1 & 1.2$\pm$0.0\\ 
	  &          &      &       &      & - & L   & 2.6$\pm$0.6 & 2.0$\pm$0.7 & 0.6$\pm$0.2 & 1.6$\pm$0.1\\
	\hline
	5 & 20110930 & 26.7 & 42.0 & Pre & + & T  & 1.8$\pm$0.6 & 1.9$\pm$0.7 & 0.3$\pm$0.1  & 0.9$\pm$0.1 \\
	  &          &      &       &    & - & L  & 2.0$\pm$0.7 & 0.8$\pm$0.5 & 0.2$\pm$0.0 & 0.3$\pm$0.0\\
	\hline
	8 & 20120614 & 96.8 & 152.1 & Pre & + & L  & 2.4$\pm$0.6 & 0.7$\pm$0.3 & 0.4$\pm$0.1 & 0.6$\pm$0.0 \\
	  &          &      &       &     & - & T  & 1.6$\pm$0.6 & 1.7$\pm$0.7 & 1.0$\pm$0.0 & 1.9$\pm$0.0\\
	\hline
	13 & 20130812 & 53.0 & 83.3 & Post & + & L  & 2.3$\pm$0.6 & 1.8$\pm$0.7 & 0.4$\pm$0.2 & 1.0$\pm$0.1\\
	   &          &      &      &      & - & T  & 1.6$\pm$0.6 & 0.9$\pm$0.6 & 0.2$\pm$0.1  & 0.6$\pm$0.1\\
	\hline
	16 & 20131013 & 47.9 & 75.3 & Post & + & L  &  1.8$\pm$0.6 & 1.7$\pm$0.7 & 0.4$\pm$0.3 & 1.0$\pm$0.1\\
	   &          &      &      &      & - & T  & 1.4$\pm$0.6 & 1.6$\pm$0.6  & 0.7$\pm$0.1  & 1.5$\pm$0.1\\
	\hline
	21 & 20140825 & 58.3 & 91.5 & Pre &  + & T  & 1.7$\pm$0.6 & 1.9$\pm$0.7 & 0.8$\pm$0.2 & 1.7$\pm$0.1\\
	   &          &      &      &     &  - & L  & 2.2$\pm$0.5 & 1.5$\pm$0.6 & 0.5$\pm$0.2  & 1.1$\pm$0.0\\
	\hline
	23 & 20140908 & 64.2 & 100.9 & Pre & + & L  & 1.8$\pm$0.5 & 1.0$\pm$0.6 & 0.4$\pm$0.0 & 0.5$\pm$0.0\\
	   &          &      &       &     & - & T  & 1.1$\pm$0.4 & 1.7$\pm$0.7 & 0.1$\pm$0.0 & 0.8$\pm$0.0\\
	\hline
	27 & 20150622 & 97.1 & 152.5 & Post & + & T & 1.8$\pm$0.6 & 1.8$\pm$0.7 & 0.7$\pm$0.1 & 1.5$\pm$0.0\\
	   &          &      &       &      & - & L & 1.4$\pm$0.6 & 2.0$\pm$0.7  & 0.4$\pm$0.2  & 1.8$\pm$0.1\\
	\enddata
	\tablecomments{ Table~\ref{tab4} shows the estimated value of magnetic twist from three methods (Tw1,Tw2,Tw3). The superscript '+'/'-' represent the twist calculated in the positive/negative footpoints respectively, and 'L/T' represent the twist calculated in the leading/trailing footpoints. The subscript 'net'/'DC' mean that we calculate Tw3 using $I_{z}^{net}$ or $I_{z}^{DC}$. The errors for Tw1 and Tw2 are the standard deviation of $\tau$, while the error for Tw3 comes from $\delta{\Phi}$ and $\delta{I}$.}
\end{deluxetable*}

\clearpage

\section{Summary \& discussion} \label{sec:dis}
In this study, we investigate 28 eruptive events that exhibit obvious conjugate coronal dimmings. The AIA/SDO observations and HMI/SDO vector magnetograms are used to identify the footprints of erupting MFRs. Our results show two MFR categories, with or without significant net electric current. The two MFR populations have distinctive observational characteristics, implying different pre-eruptive magnetic field and evolution of these MFRs toward eruptions. In the following sections, we will first summarize what we have found and further compare our results with the existing flux-rope models.

Here we summarize what we have learned from our quantitative measurements of pre-eruptive magnetic properties of 28 erupting MFRs, whose feet are well identified by conjugate coronal dimmings:
~\\1.In our sample, we find 9 events with pre-eruption conjugate dimmings, while the rest with post-eruption conjugate dimmings. All pre-eruption dimming events are accompanied by coronal structures considered to be plasma signatures of MFRs (e.g. sigmoids, filaments, expanding coronal structures). The pre-eruption dimmings evolve simultaneously with the MFR-like structures until they erupt as CMEs. Most post-eruption dimmings appear several minutes or dozens of minutes after the onset of flares. 
~\\2.Quantitative measurements of electric currents in 28 MFRs' conjugate feet show that only 8 of them carry significant non-neutralized currents ($R_{z}>2.0$), that the rest might carry neutralized currents. The difference of magnetic flux and direct current between two MFR categories is not statistically significant (Figure~\ref{fig10}).
~\\3.The MFRs carrying non-neutralized currents exhibit the asymmetric electric current and magnetic twist at their feet. For most MFRs, electric currents ($I_{net}$ and $I_{DC}$) in one foot are almost 2 to 4 times larger than the other one. For the average twist, the imbalance reduces to a factor of two. The Rz can be asymmetric, too, with non-neutralized current at one foot and neutralized current at the other. In particular, electric current may be concentrated in the form of a ribbon at one foot, but rather diffuse at the other (Figure~\ref{fig12}). 

\subsection{The pre-eruptive magnetic field: flux-rope models or others}
There has been a long-standing debate on the pre-eruptive magnetic field of erupting MFRs. In our study, combining evolution of pre-eruption dimmings and electric currents within dimming regions, we found signatures of different pre-eruptive magnetic fields. Different evolution of dimming fluxes are observed in the pre-eruption dimming events with or without net currents. For the pre-eruption dimming events with non-neutralized currents (see Figure~\ref{fig1}(a)(d)), the dimming regions all undergo a rapidly expansion followed by quasi-static evolution stage before the eruption. The dimming fluxes will reach the maximum before the eruption. For example, in the 20120614 event, the expanding coronal structure observed by STEREO-B experienced five-hours slow rising at speed of about 2 to 5 km s$^{-1}$ (see Figure~\ref{fig7}) before the eruption. The expanding coronal structure finally evolved as a halo CME, which was identified as an MC \citep{james2017disc} two days later when it passed through WIND spacecraft. Its footpoints as identified by pre-eruption dimmings carry non-neutralized currents ($R_{z}>2.0$), implying a pre-existing current-carrying MFR. The pre-existing MFR was found by \cite{james2018observationally} using NLFFF extrapolation (see Figure 3 in their paper). Therefore, we suggest that pre-eruption dimmings and non-neutralized currents are consistent with the pre-existing current-carrying MFR.
	
But for the rest pre-eruption dimming events with neutralized currents, the dimming regions experience a slow expansion followed by a stable stage (see Figure~\ref{fig1}(b)(c)), which is similar to the evolution of some post-eruption dimmings. For these pre-eruption dimmings, the dimming fluxes start to rise before the eruption, but reach the maximum during the eruption. \cite{zhou2017toward} had studied the 20110621 event in detail. A pre-eruption sigmoidal structure was observed in the EUV hot channel and shared the same location with a filament (see Figure 2 in their paper). The conjugate dimmings detected in our study are located in the two ends of the filament-sigmoid structure. The $R_{z}$ within the two identified footpoints are around 1.0, implying a neutralized flux-rope model.

The situation is more complicated for the post-eruption dimming events. Only 4 of 19 post-eruption dimming events carry non-neutralized currents (Table~\ref{tab3}). For these four events, high electric current densities of different sign distribute on the two sides of the PIL and their identified footpoints anchor at two ends of current ribbons (see Figure~\ref{fig11}), conforming to the 3D extension of standard flare model \citep{janvier2014electric}. For example, for the 20130812 event, \cite{liu2016structure} had investigated the host active region 11817 during 2013 August 10 to 12. Aided by NLFFF method, they identified a Double-decker MFR system one day before the 20130812 event. \cite{awasthi2018pre-eruptive} also applied the NLFFF extrapolation method to the 20150622 event and found a pre-existing multi-flux-rope system (see the Figure 1 in their paper). These studies further suggest that the post-eruption dimmings with non-neutralized currents may be also associated with pre-existing MFRs. But for the other post-eruption dimming events, the evolution of dimmings reflect the formation process of erupting MFRs. The onset of the dimming typically lags the onset of the flare by several minutes or even several tens of minutes. \cite{wang2017buildup} had investigated the 20151104 event and found that the main body of a highly twisted MFR was formed during the eruption via magnetic reconnection, which was associated with the development of dimmings inside a pair of closed hooked ribbons.

\subsection{The role of magnetic reconnection in the dynamic evolution of conjugate dimmings} \label{sec:dis-dim}
For both pre-eruption and post-eruption dimming events, both flare ribbons and dimmings are dynamically evolving, as a result, the accumulative ribbons and dimming regions overlap each other (see Figure~\ref{fig3} and Figure~\ref{fig4}). For pre-eruption dimmings, flare ribbons will erode part of dimming areas. For the 20110930 and 20120614 event, more than half of dimming areas disappear after the flares (see Figure~\ref{fig8} (a) and (d)). The dimming areas can also extend in area after the flare, e.g. the 20130830 event (Figure~\ref{fig8} (b)), or be almost unchanged after the flare, e.g. the 20110621 event (Figure~\ref{fig8} (c)). The loss of pre-eruption dimming areas may reflect internal changes of the pre-existing MFRs, suggesting that the pre-existing MFRs may undergo internal reconnections due to the presence of QSLs inside the rope (e.g., \citealt{awasthi2018pre-eruptive}), which are termed 'rr-rf' reconnections in \cite{aulanier2019drifting}, with `r' referring to the rope field and 'f' to post-flare loops. For most post-eruption dimming events, dimming regions are partially swept by flare ribbons (see Figure~\ref{fig4}), implying MFRs may be built up via reconnection during the eruptions. The complexity of the role of 3D reconnection in buildup of MFRs requires further investigation beyond the scope of this study.

In our sample, morphological evolution of both pre-eruption and post-eruption dimmings are relatively stable. But we also find several events with rapid drifting dimmings, for example, the 20120712 event. This event was also well investigated by many previous studies (e.g. \citealt{dudik2014slipping,cheng2014formation,cheng2016characteristics}). Apparent slipping motions of both flare and erupting loops were observed in this event (see Figure 6,8,10 in \citealt{dudik2014slipping}). \cite{dudik2014slipping} further indicated that the slipping motions fed	the MFR with twisted field lines surrounding its core, leading to the MFR expansion. As a result, the dimmings drift following the slipping motions. The drifting of dimmings may reflect the drifting of the MFR footpoints. \cite{aulanier2019drifting} indicated that a series of coronal reconnections can change the foot-point area of flux rope. They also suggested two new reconnection terminologies: aa-rf and ar-rf reconnections (see Figure 4, Figure 5, and Figure 6 in their paper), with 'a' referring to the arcade field. Distinctive morphological evolution of conjugate dimmings may suggest different processes of reconnection.

\subsection{The non-neutralized current inside the MFR} \label{sec:dis-cur}

Identification of the MFR's feet is very challenging. Our method outlines possible areas of footpoints by detecting conjugate coronal dimmings. It is important to keep in mind that conjugate dimmings can map the footpoints of erupting MFRs but may not cover the whole region of the footpoints. On the other hand, our results are subject to the accuracy of identification, which has been explained in Appendix~\ref{app:A}.

Distinctive distributions of electric current at the MFR feet may imply different internal structures of flux-ropes. For the MFRs with non-neutralized currents, high electric current densities can concentrate on the center of the footpoint (e.g. the 20130812 event), or the boundary of the footpoint (e.g. the 20131012 event). Furthermore, high electric current densities can develop into smooth continuous ribbons, e.g a spiral-like ribbon (the 20110930 event) and a semiarc-like ribbon (the 20120614 event). But it is hard to tell the distribution of electric current within the MFRs with neutralized currents, when opposite signs of current densities are mixed and randomly distributed. In our sample, most footpoints associated with neutralized currents are located in the relatively weak field. The main problem here is uncertainties of vertical electric current, especially in the weak field, due to large uncertainties of transverse field. It is important to keep in mind that our results only give a rough estimate of the degree of current neutralization of these MFRs. Intruments with higher resolution and sensitivity (e.g the Daniel K. Inouye Solar Telescope) are required to reveal more detailed current distributions within the MFRs.

Previous numerical simulations indicated that photospheric flows may play a significant role in the development of net current in the solar active regions (e.g. \citealt{torok2014distribution,dalmasse2015origin}). In the MHD simulation of \cite{torok2014distribution}, the buildup of net current inside the MFR occurred when strong shear developed along the PIL. Indeed, shear flows at speed of 1 km s$^{-1}$ are observed along the PILs for the three events in Figure~\ref{fig11}. Particularly, for the 20120614 event, the semiarc-like current ribbon observed in the FP- (Figure~\ref{fig12} (b)) shares the same location with shear flows. A parametric MHD simulation from \cite{dalmasse2015origin} also confirmed that both shear motions and twisting motions below the MFR could inject net current into its feet. Similarly, for the 20110930 event, a spiral-like current ribbon is observed in the footpoint of the MFR, which is co-spatial with the rotational sunspot (Figure~\ref{fig12} (a)). Different from their simulations, however, the observed twisting motion only occurs in one foot of the studied MFR, which may explain the asymmetric electric current and magnetic twist distribution. More evidences are required to verify the origin of net currents.

~\\
In conclusion, we investigate pre-eruptive magnetic properties at footpoints of 28 MFRs using HMI vector magnetogram. Our statistical study indicates that about 28\% (8 out of 28) of the MFRs carry significant non-neutralized currents ($R_{z}>$2.0) and half of them are associated with pre-eruption dimmings at the footpoints, suggesting that such MFRs are most likely formed prior to eruption. The distributions of electric current and magnetic twist at the MFR footpoints as well as the asymmetry of the distributions may help us to diagnose the internal structures of the MFRs and to further may shed light on their formation mechanism.

\begin{acknowledgments}
We thank two unknown referees for very helpful comments. This work was supported by National Natural Science Foundation of China (NSFC 12003032, 11925302, 4218810), the China Postdoctoral Science Foundation (Grant No. 2019TQ0313), the Strategic Priority Program of the Chinese Academy of Sciences (XDB41000000), and the Fundamental Research Funds for the Central Universities.
\end{acknowledgments}

\bibliographystyle{aasjournal} 
\bibliography{manuscript}

\begin{thebibliography}{}
\expandafter\ifx\csname natexlab\endcsname\relax\def\natexlab#1{#1}\fi
\providecommand{\url}[1]{\href{#1}{#1}}
\providecommand{\dodoi}[1]{doi:~\href{http://doi.org/#1}{\nolinkurl{#1}}}
\providecommand{\doeprint}[1]{\href{http://ascl.net/#1}{\nolinkurl{http://ascl.net/#1}}}
\providecommand{\doarXiv}[1]{\href{https://arxiv.org/abs/#1}{\nolinkurl{https://arxiv.org/abs/#1}}}

\bibitem[{{Aulanier} \& {Dud{\'\i}k}(2019)}]{aulanier2019drifting}
{Aulanier}, G., \& {Dud{\'\i}k}, J. 2019, \aap, 621, A72,
  \dodoi{10.1051/0004-6361/201834221}

\bibitem[{{Avallone} \& {Sun}(2020)}]{avallone2020electric}
{Avallone}, E.~A., \& {Sun}, X. 2020, \apj, 893, 123,
  \dodoi{10.3847/1538-4357/ab7afa}

\bibitem[{{Awasthi} {et~al.}(2018){Awasthi}, {Liu}, {Wang}, {Wang}, \&
  {Shen}}]{awasthi2018pre-eruptive}
{Awasthi}, A.~K., {Liu}, R., {Wang}, H., {Wang}, Y., \& {Shen}, C. 2018, \apj,
  857, 124, \dodoi{10.3847/1538-4357/aab7fb}

\bibitem[{{Barczynski} {et~al.}(2020){Barczynski}, {Aulanier}, {Janvier},
  {Schmieder}, \& {Masson}}]{barczynski2020electric}
{Barczynski}, K., {Aulanier}, G., {Janvier}, M., {Schmieder}, B., \& {Masson},
  S. 2020, \apj, 895, 18, \dodoi{10.3847/1538-4357/ab893d}

\bibitem[{Bobra {et~al.}(2014)Bobra, Sun, Hoeksema, Turmon, Liu, Hayashi,
  Barnes, \& Leka}]{bobra2014helioseismic}
Bobra, M.~G., Sun, X., Hoeksema, J.~T., {et~al.} 2014, Solar Physics, 289, 3549

\bibitem[{Cheng \& Qiu(2016)}]{cheng2016nature}
Cheng, J., \& Qiu, J. 2016, The Astrophysical Journal, 825, 37

\bibitem[{Cheng \& Ding(2016)}]{cheng2016characteristics}
Cheng, X., \& Ding, M. 2016, The Astrophysical Journal Supplement Series, 225,
  16

\bibitem[{{Cheng} {et~al.}(2014){Cheng}, {Ding}, {Zhang}, {Sun}, {Guo}, {Wang},
  {Kliem}, \& {Deng}}]{cheng2014formation}
{Cheng}, X., {Ding}, M.~D., {Zhang}, J., {et~al.} 2014, \apj, 789, 93,
  \dodoi{10.1088/0004-637X/789/2/93}

\bibitem[{{Cheung} \& {Isobe}(2014)}]{cheung2014flux}
{Cheung}, M. C.~M., \& {Isobe}, H. 2014, Living Reviews in Solar Physics, 11,
  3, \dodoi{10.12942/lrsp-2014-3}

\bibitem[{Dalmasse {et~al.}(2015)Dalmasse, Aulanier, D{\'e}moulin, Kliem,
  T{\'o}r{\'o}k, \& Pariat}]{dalmasse2015origin}
Dalmasse, K., Aulanier, G., D{\'e}moulin, P., {et~al.} 2015, \apj

\bibitem[{{D{\'e}moulin} \& {Aulanier}(2010)}]{demoulin2010criteria}
{D{\'e}moulin}, P., \& {Aulanier}, G. 2010, \apj, 718, 1388,
  \dodoi{10.1088/0004-637X/718/2/1388}

\bibitem[{{Dissauer} {et~al.}(2019){Dissauer}, {Veronig}, {Temmer}, \&
  {Podladchikova}}]{dissauer2019statistics}
{Dissauer}, K., {Veronig}, A.~M., {Temmer}, M., \& {Podladchikova}, T. 2019,
  \apj, 874, 123, \dodoi{10.3847/1538-4357/ab0962}

\bibitem[{{Dissauer} {et~al.}(2018{\natexlab{a}}){Dissauer}, {Veronig},
  {Temmer}, {Podladchikova}, \& {Vanninathan}}]{dissauer2018detection}
{Dissauer}, K., {Veronig}, A.~M., {Temmer}, M., {Podladchikova}, T., \&
  {Vanninathan}, K. 2018{\natexlab{a}}, \apj, 855, 137,
  \dodoi{10.3847/1538-4357/aaadb5}

\bibitem[{{Dissauer} {et~al.}(2018{\natexlab{b}}){Dissauer}, {Veronig},
  {Temmer}, {Podladchikova}, \& {Vanninathan}}]{dissauer2018statistics}
---. 2018{\natexlab{b}}, \apj, 863, 169, \dodoi{10.3847/1538-4357/aad3c6}

\bibitem[{{Dud{\'\i}k} {et~al.}(2014){Dud{\'\i}k}, {Janvier}, {Aulanier}, {Del
  Zanna}, {Karlick{\'y}}, {Mason}, \& {Schmieder}}]{dudik2014slipping}
{Dud{\'\i}k}, J., {Janvier}, M., {Aulanier}, G., {et~al.} 2014, \apj, 784, 144,
  \dodoi{10.1088/0004-637X/784/2/144}

\bibitem[{Fan(2009)}]{fan2009emergence}
Fan, Y. 2009, The Astrophysical Journal, 697, 1529

\bibitem[{Fan \& Gibson(2004)}]{fan2004numerical}
Fan, Y., \& Gibson, S. 2004, The Astrophysical Journal, 609, 1123

\bibitem[{{Georgoulis} {et~al.}(2012){Georgoulis}, {Titov}, \&
  {Miki{\'c}}}]{georgoulis2012non}
{Georgoulis}, M.~K., {Titov}, V.~S., \& {Miki{\'c}}, Z. 2012, \apj, 761, 61,
  \dodoi{10.1088/0004-637X/761/1/61}

\bibitem[{Gosling(1990)}]{gosling1990coronal}
Gosling, J.~T. 1990, Physics of magnetic flux ropes, 343

\bibitem[{Harra {et~al.}(2007)Harra, Hara, Imada, Young, Williams, Sterling,
  Korendyke, \& Attrill}]{harra2007coronal}
Harra, L.~K., Hara, H., Imada, S., {et~al.} 2007, Publications of the
  Astronomical Society of Japan, 59, S801

\bibitem[{Harra \& Sterling(2001)}]{harra2001material}
Harra, L.~K., \& Sterling, A.~C. 2001, The Astrophysical Journal Letters, 561,
  L215

\bibitem[{Harrison \& Lyons(2000)}]{harrison2000spectroscopic}
Harrison, R., \& Lyons, M. 2000, Astronomy and Astrophysics, 358, 1097

\bibitem[{Harvey \& Recely(2002)}]{harvey2002polar}
Harvey, K.~L., \& Recely, F. 2002, Solar Physics, 211, 31

\bibitem[{Hoeksema {et~al.}(2014)Hoeksema, Liu, Hayashi, Sun, Schou, Couvidat,
  Norton, Bobra, Centeno, Leka, {et~al.}}]{hoeksema2014helioseismic}
Hoeksema, J.~T., Liu, Y., Hayashi, K., {et~al.} 2014, Solar Physics, 289, 3483

\bibitem[{Hood \& Priest(1979)}]{hood1979kink}
Hood, A.~W., \& Priest, E. 1979, Solar Physics, 64, 303

\bibitem[{Hu {et~al.}(2014)Hu, Qiu, Dasgupta, Khare, \&
  Webb}]{hu2014structures}
Hu, Q., Qiu, J., Dasgupta, B., Khare, A., \& Webb, G. 2014, The Astrophysical
  Journal, 793, 53

\bibitem[{James {et~al.}(2017)James, Green, Palmerio, Valori, Reid, Baker,
  Brooks, van Driel-Gesztelyi, \& Kilpua}]{james2017disc}
James, A., Green, L., Palmerio, E., {et~al.} 2017, Solar Physics, 292, 71

\bibitem[{{James} {et~al.}(2018){James}, {Valori}, {Green}, {Liu}, {Cheung},
  {Guo}, \& {van Driel-Gesztelyi}}]{james2018observationally}
{James}, A.~W., {Valori}, G., {Green}, L.~M., {et~al.} 2018, \apjl, 855, L16,
  \dodoi{10.3847/2041-8213/aab15d}

\bibitem[{{Janvier} {et~al.}(2014){Janvier}, {Aulanier}, {Bommier},
  {Schmieder}, {D{\'e}moulin}, \& {Pariat}}]{janvier2014electric}
{Janvier}, M., {Aulanier}, G., {Bommier}, V., {et~al.} 2014, \apj, 788, 60,
  \dodoi{10.1088/0004-637X/788/1/60}

\bibitem[{Kaiser {et~al.}(2008)Kaiser, Kucera, Davila, Cyr, Guhathakurta, \&
  Christian}]{kaiser2008stereo}
Kaiser, M.~L., Kucera, T., Davila, J., {et~al.} 2008, Space Science Reviews,
  136, 5

\bibitem[{{Kazachenko} {et~al.}(2017){Kazachenko}, {Lynch}, {Welsch}, \&
  {Sun}}]{kazachenko2017database}
{Kazachenko}, M.~D., {Lynch}, B.~J., {Welsch}, B.~T., \& {Sun}, X. 2017, \apj,
  845, 49, \dodoi{10.3847/1538-4357/aa7ed6}

\bibitem[{{Kontogiannis} {et~al.}(2019){Kontogiannis}, {Georgoulis}, {Guerra},
  {Park}, \& {Bloomfield}}]{kontogiannis2019which}
{Kontogiannis}, I., {Georgoulis}, M.~K., {Guerra}, J.~A., {Park}, S.-H., \&
  {Bloomfield}, D.~S. 2019, \solphys, 294, 130,
  \dodoi{10.1007/s11207-019-1523-6}

\bibitem[{{Leka} {et~al.}(2009){Leka}, {Barnes}, {Crouch}, {Metcalf}, {Gary},
  {Jing}, \& {Liu}}]{leka2009resolving}
{Leka}, K.~D., {Barnes}, G., {Crouch}, A.~D., {et~al.} 2009, \solphys, 260, 83,
  \dodoi{10.1007/s11207-009-9440-8}

\bibitem[{{Leka} {et~al.}(1996){Leka}, {Canfield}, {McClymont}, \& {van
  Driel-Gesztelyi}}]{leka1996evidence}
{Leka}, K.~D., {Canfield}, R.~C., {McClymont}, A.~N., \& {van Driel-Gesztelyi},
  L. 1996, \apj, 462, 547, \dodoi{10.1086/177171}

\bibitem[{Lemen {et~al.}(2012)Lemen, Akin, Boerner, Chou, Drake, Duncan,
  Edwards, Friedlaender, Heyman, Hurlburt, {et~al.}}]{lemen2012atmospheric}
Lemen, J.~R., Akin, D.~J., Boerner, P.~F., {et~al.} 2012, Solar physics, 275,
  17

\bibitem[{Liu {et~al.}(2016)Liu, Kliem, Titov, Chen, Wang, Wang, Liu, Xu, \&
  Wiegelmann}]{liu2016structure}
Liu, R., Kliem, B., Titov, V.~S., {et~al.} 2016, The Astrophysical Journal,
  818, 148

\bibitem[{Liu {et~al.}(2017)Liu, Sun, T{\"o}r{\"o}k, Titov, \&
  Leake}]{liu2017electric}
Liu, Y., Sun, X., T{\"o}r{\"o}k, T., Titov, V.~S., \& Leake, J.~E. 2017, The
  Astrophysical Journal Letters, 846, L6

\bibitem[{{Longcope} \& {Beveridge}(2007)}]{longcope2007quantitative}
{Longcope}, D.~W., \& {Beveridge}, C. 2007, \apj, 669, 621,
  \dodoi{10.1086/521521}

\bibitem[{{Longcope} \& {Welsch}(2000)}]{longcope2000model}
{Longcope}, D.~W., \& {Welsch}, B.~T. 2000, \apj, 545, 1089,
  \dodoi{10.1086/317846}

\bibitem[{{Melrose}(1991)}]{melrose1991neutralized}
{Melrose}, D.~B. 1991, \apj, 381, 306, \dodoi{10.1086/170652}

\bibitem[{{Melrose}(1995)}]{melrose1995current}
---. 1995, \apj, 451, 391, \dodoi{10.1086/176228}

\bibitem[{{Metcalf}(1994)}]{metcalf1994resolving}
{Metcalf}, T.~R. 1994, \solphys, 155, 235, \dodoi{10.1007/BF00680593}

\bibitem[{{Metcalf} {et~al.}(2006){Metcalf}, {Leka}, {Barnes}, {Lites},
  {Georgoulis}, {Pevtsov}, {Balasubramaniam}, {Gary}, {Jing}, {Li}, {Liu},
  {Wang}, {Abramenko}, {Yurchyshyn}, \& {Moon}}]{metcalf2006overview}
{Metcalf}, T.~R., {Leka}, K.~D., {Barnes}, G., {et~al.} 2006, \solphys, 237,
  267, \dodoi{10.1007/s11207-006-0170-x}

\bibitem[{Moore {et~al.}(2001)Moore, Sterling, Hudson, \&
  Lemen}]{moore2001onset}
Moore, R.~L., Sterling, A.~C., Hudson, H.~S., \& Lemen, J.~R. 2001, The
  Astrophysical Journal, 552, 833

\bibitem[{{Parker}(1996)}]{parker1996inferring}
{Parker}, E.~N. 1996, \apj, 471, 485, \dodoi{10.1086/177983}

\bibitem[{Pesnell {et~al.}(2011)Pesnell, Thompson, \&
  Chamberlin}]{pesnell2011solar}
Pesnell, W.~D., Thompson, B.~J., \& Chamberlin, P. 2011, in The Solar Dynamics
  Observatory (Springer), 3--15

\bibitem[{Qiu \& Cheng(2017)}]{qiu2017gradual}
Qiu, J., \& Cheng, J. 2017, The Astrophysical Journal Letters, 838, L6

\bibitem[{Qiu {et~al.}(2007)Qiu, Hu, Howard, \& Yurchyshyn}]{qiu2007magnetic}
Qiu, J., Hu, Q., Howard, T.~A., \& Yurchyshyn, V.~B. 2007, The Astrophysical
  Journal, 659, 758

\bibitem[{Scholl \& Habbal(2008)}]{scholl2008automatic}
Scholl, I.~F., \& Habbal, S.~R. 2008, Solar Physics, 248, 425

\bibitem[{Sun \& Cheung(2020)}]{sun2020non}
Sun, X., \& Cheung, M. C.~M. 2020, Solar physics

\bibitem[{Titov \& D{\'e}moulin(1999)}]{titov1999basic}
Titov, V., \& D{\'e}moulin, P. 1999, Astronomy and Astrophysics, 351, 707

\bibitem[{T{\"o}r{\"o}k {et~al.}(2004)T{\"o}r{\"o}k, Kliem, \&
  Titov}]{torok2004ideal}
T{\"o}r{\"o}k, T., Kliem, B., \& Titov, V. 2004, Astronomy \& Astrophysics,
  413, L27

\bibitem[{{T{\"o}r{\"o}k} {et~al.}(2014){T{\"o}r{\"o}k}, {Leake}, {Titov},
  {Archontis}, {Miki{\'c}}, {Linton}, {Dalmasse}, {Aulanier}, \&
  {Kliem}}]{torok2014distribution}
{T{\"o}r{\"o}k}, T., {Leake}, J.~E., {Titov}, V.~S., {et~al.} 2014, \apjl, 782,
  L10, \dodoi{10.1088/2041-8205/782/1/L10}

\bibitem[{{Vemareddy}(2019)}]{vemareddy2019degree}
{Vemareddy}, P. 2019, \mnras, 486, 4936, \dodoi{10.1093/mnras/stz1020}

\bibitem[{Wang {et~al.}(2017)Wang, Liu, Wang, Hu, Shen, Jiang, \&
  Zhu}]{wang2017buildup}
Wang, W., Liu, R., Wang, Y., {et~al.} 2017, Nature communications, 8, 1330

\bibitem[{Wang {et~al.}(2019)Wang, Zhu, Qiu, Liu, Yang, \&
  Hu}]{wang2019evolution}
Wang, W., Zhu, C., Qiu, J., {et~al.} 2019, The Astrophysical Journal, 871, 25

\bibitem[{Webb {et~al.}(2000)Webb, Cliver, Crooker, St~Cyr, \&
  Thompson}]{webb2000relationship}
Webb, D., Cliver, E., Crooker, N., St~Cyr, O., \& Thompson, B. 2000, Journal of
  Geophysical Research: Space Physics, 105, 7491

\bibitem[{{Xing} {et~al.}(2020){Xing}, {Cheng}, \& {Ding}}]{xing2020evolution}
{Xing}, C., {Cheng}, X., \& {Ding}, M.~D. 2020, The Innovation, 1, 100059,
  \dodoi{10.1016/j.xinn.2020.100059}

\bibitem[{{Zhou} {et~al.}(2017){Zhou}, {Zhang}, {Wang}, {Liu}, \&
  {Chintzoglou}}]{zhou2017toward}
{Zhou}, Z., {Zhang}, J., {Wang}, Y., {Liu}, R., \& {Chintzoglou}, G. 2017,
  \apj, 851, 133, \dodoi{10.3847/1538-4357/aa9bd9}

\end{thebibliography}

\begin{figure}
	\epsscale{.95}
	\plotone{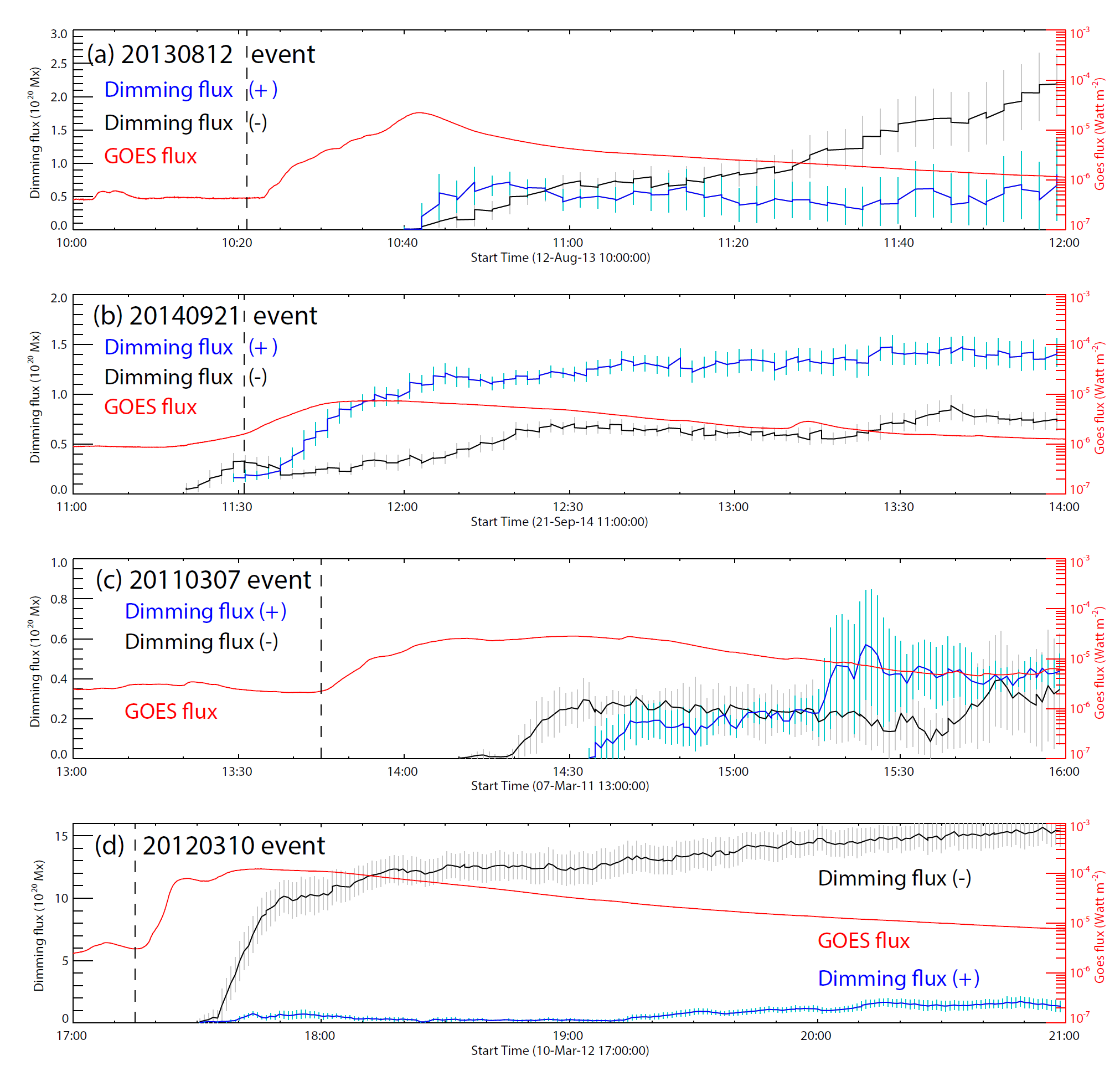}
	\caption{Evolution of post-eruption coronal dimmings for four events. All diagrams show temporal evolution of dimming fluxes (black for negative, blue for positive) and GOES flux (red). The vertical dashed line represents the onset of the flare for each event. \label{fig2}}
\end{figure}

\begin{figure}
	\epsscale{.95}
	\plotone{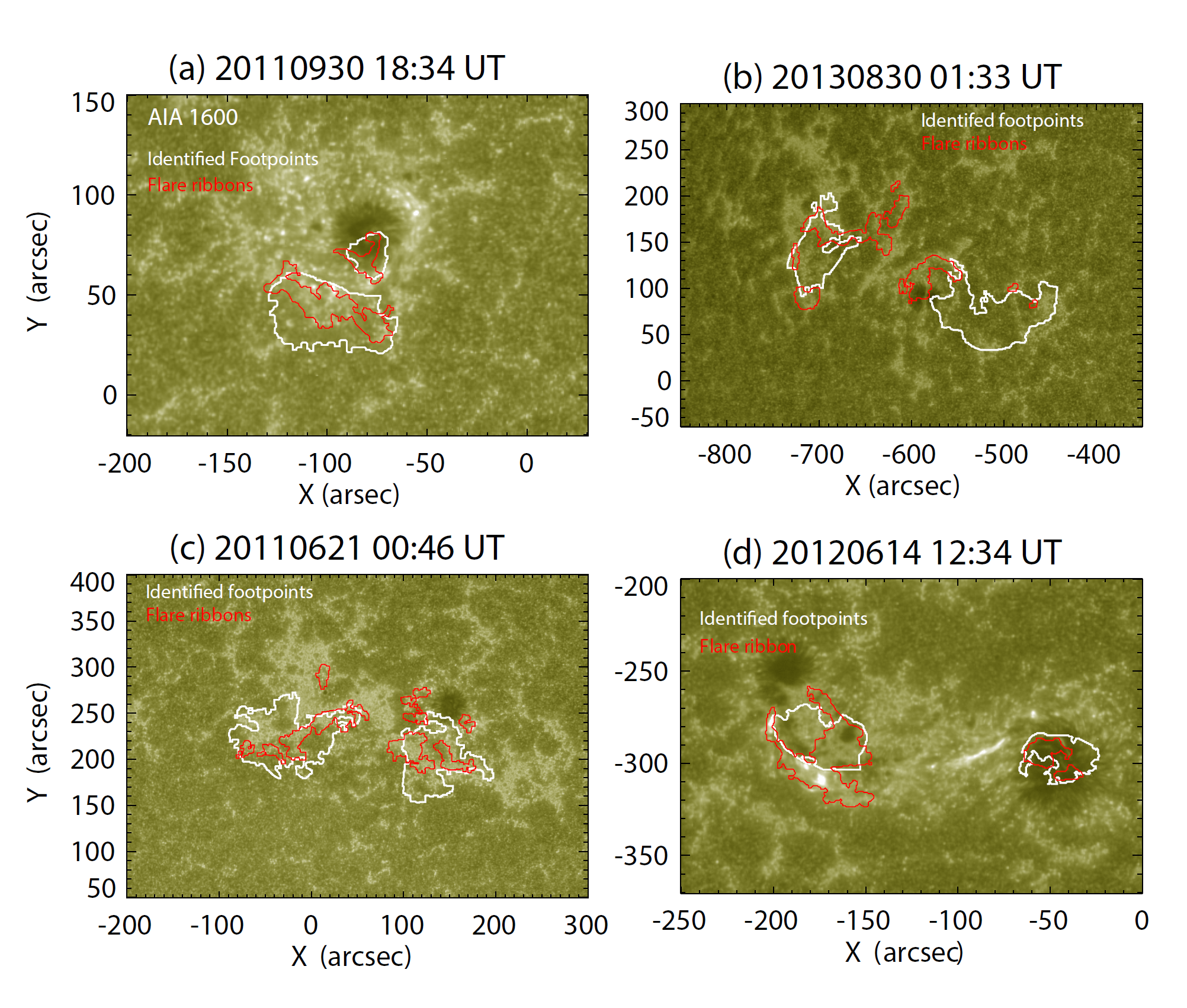}
	\caption{Pre-eruption dimming events: relative locations between identified footpoints and flare ribbons. All snapshots are from AIA 1600 channel. For each event, two contours show identified footpoints from conjugate dimmings (white) and detected flare ribbons (red). \label{fig3}}
\end{figure}

\begin{figure}
	\epsscale{.9}
	\plotone{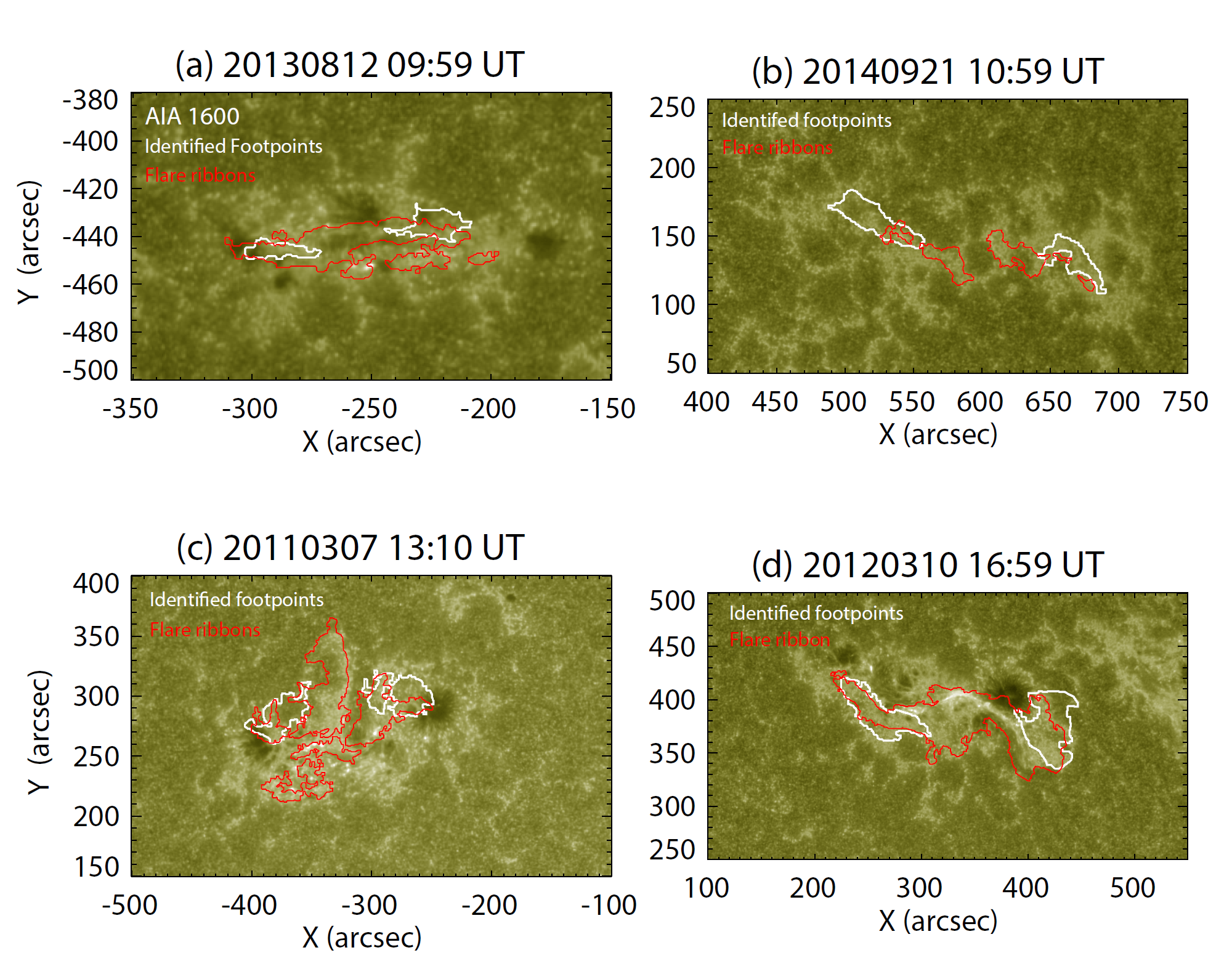}
	%	\vskip -2cm
	\caption{Post-eruption dimming events: relative locations between identified footpoints and flare ribbons. All snapshots are from AIA 1600 channel. For each event, two contours show identified footpoints from conjugate dimmings (white) and detected flare ribbons (red).\label{fig4}}
\end{figure}

\begin{figure}
	\epsscale{.95}
	\plotone{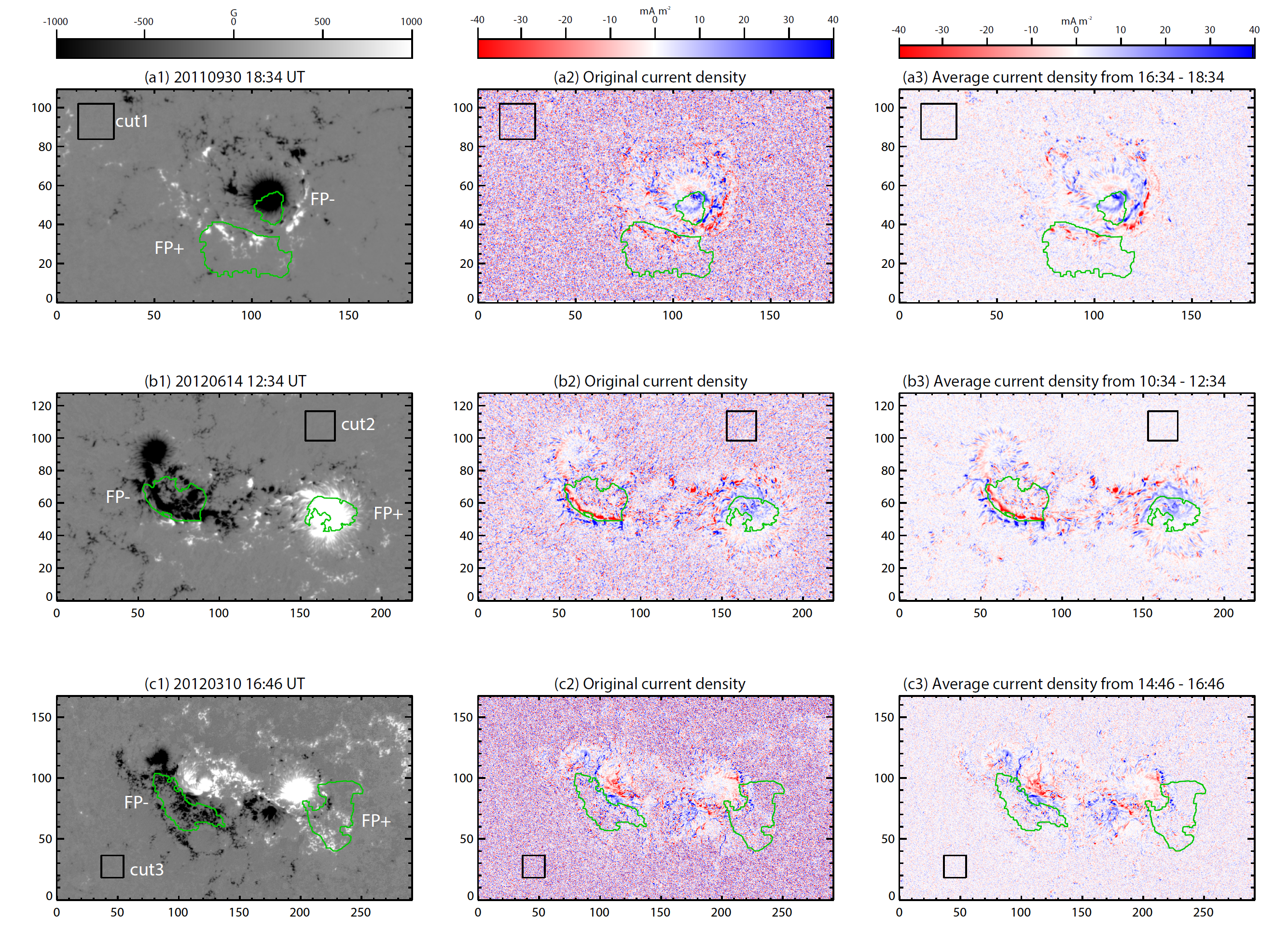}
	\caption{Examples for HMI vector magnetogram (left), original current density maps (center) and timing-averaged current density maps (right). Four rows represent four different events respectively. Green contours show the identified footpoint regions. \label{fig5}}
\end{figure}

\begin{figure}
	%\epsscale{.9}
	\plotone{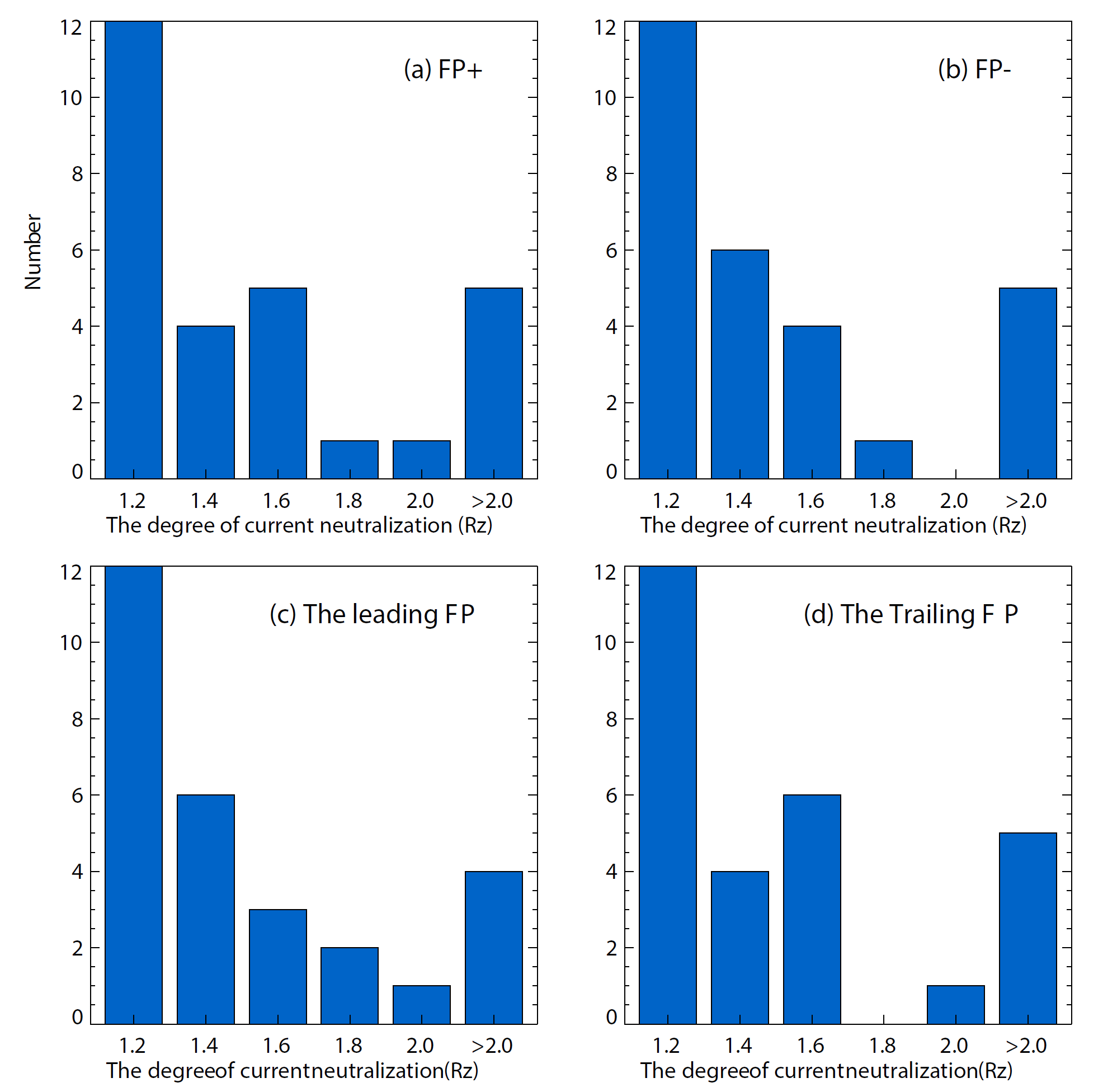}
	\caption{The histogram of degree of current neutralization $R_{z}$. (a) shows the $R_{z}$ estimated in FP+, and (b) shows the $R_{z}$ for FP-. (c) shows the $R_{z}$ in the leading FP, and (d) shows the $R_{z}$ in the trailing FP.\label{fig6}}
\end{figure}

\begin{figure}
	%\epsscale{.9}
	\plotone{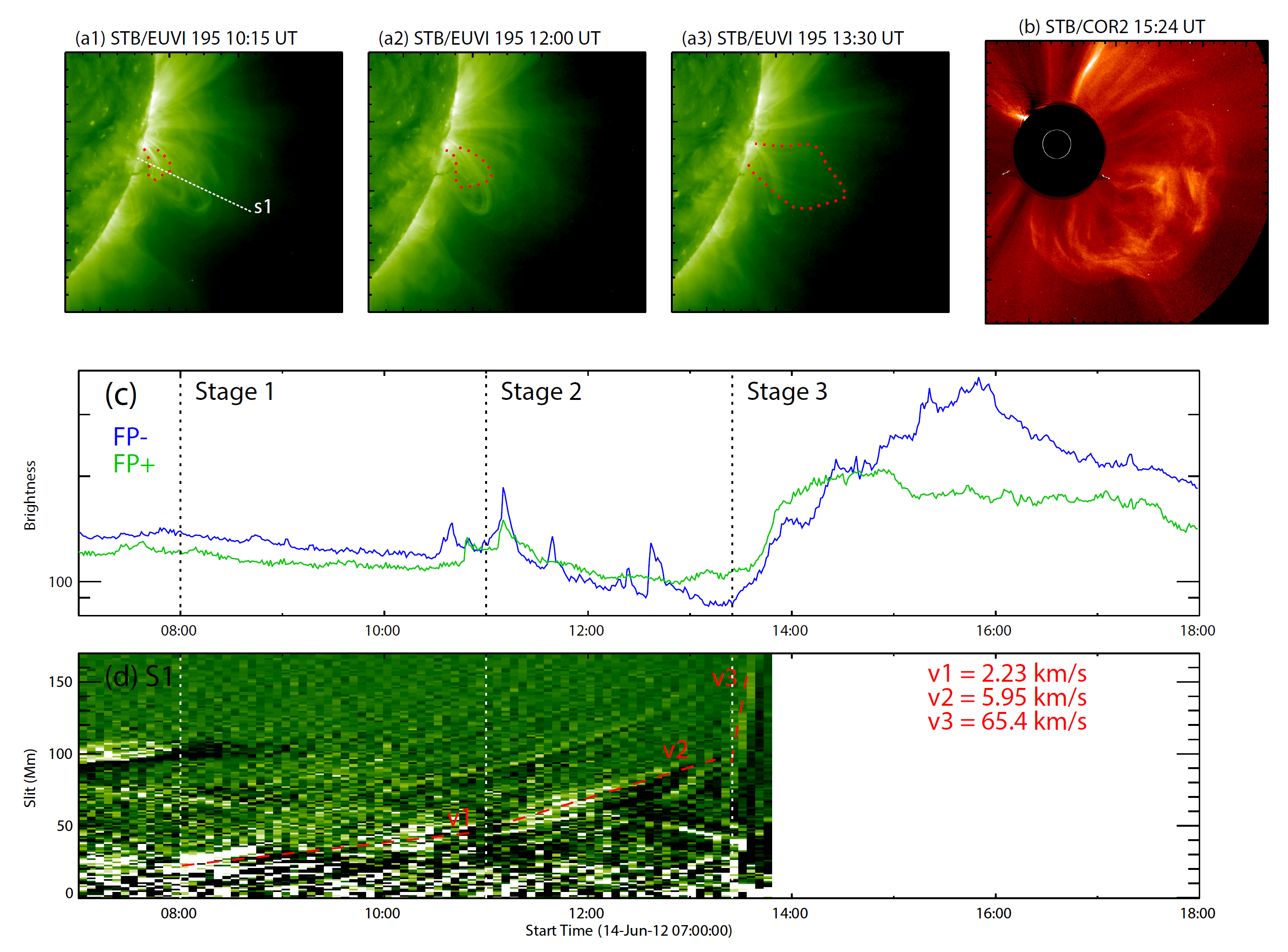}
	\caption{One typical case of pre-eruption dimmings: the 20120614 event. The panels (a1) to (a3) show the expanding coronal structure observed in 195 passband by STEREO-B/EUVI, which finally evolved as a halo CME captured by STEREO-B/COR2 (b). The diagram (c) shows the temporal profiles of brightness in AIA 304. Green and blue curves represent the average brightness in two identified footpoints (see white contours in the Figure~\ref{fig3} (d)). The diagram (d) represents the time–distance map along “s1” in (a1). Three linear fittings are indicated by red dashed lines. Three vertical dashed lines mark the beginning of three different stages. More detailed information can be found in the \cite{wang2019evolution}. \label{fig7}}
\end{figure}

\begin{figure}
	%\epsscale{.9}
	\plotone{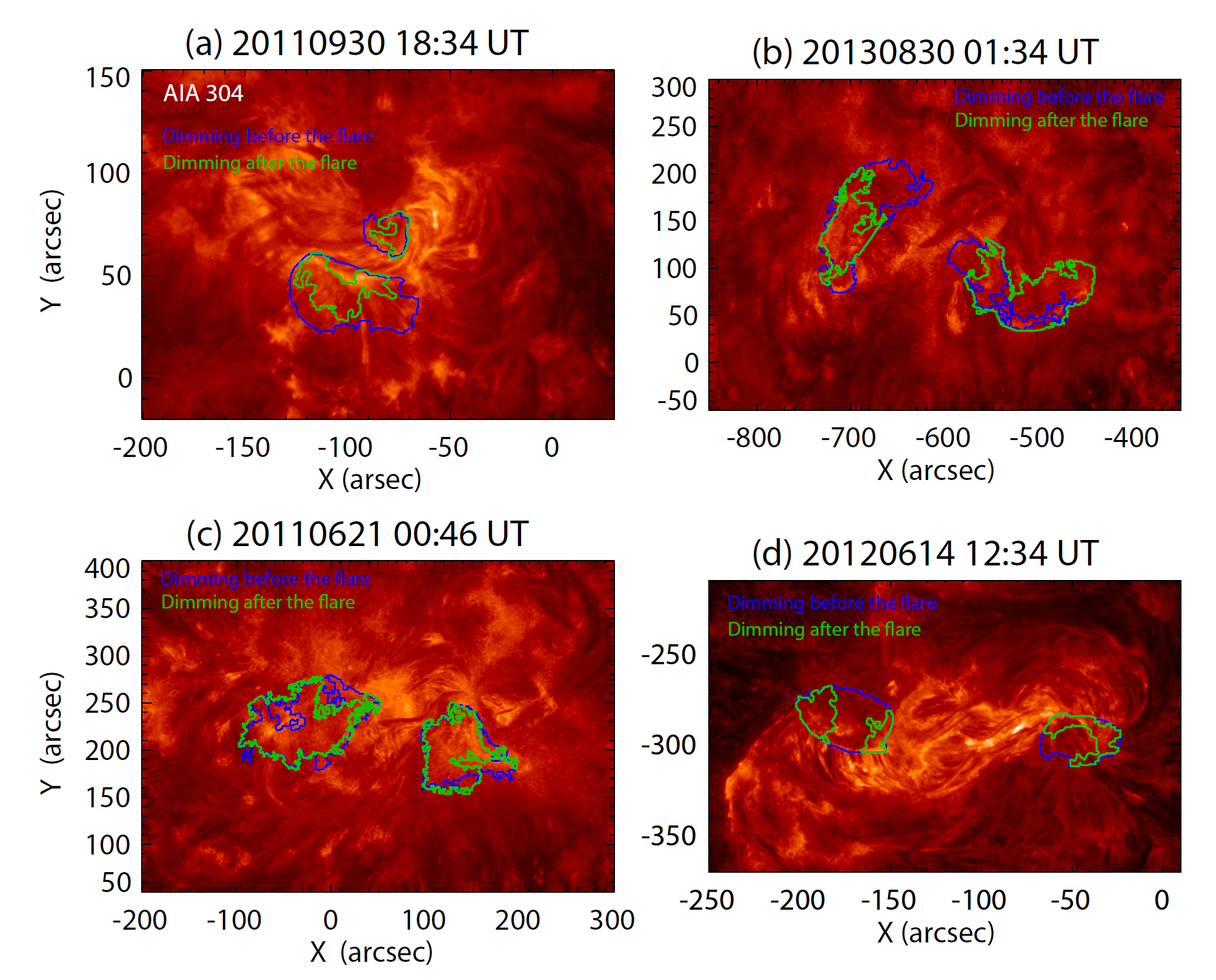}
	\caption{The observational feature of pre-eruption dimmings. All snapshots are from AIA 304 channel. For each event, two contours show the dimming regions appear before (blue) or after (green) the onset of the flare.\label{fig8}}
\end{figure}

\begin{figure}
	%\epsscale{.9}
	\plotone{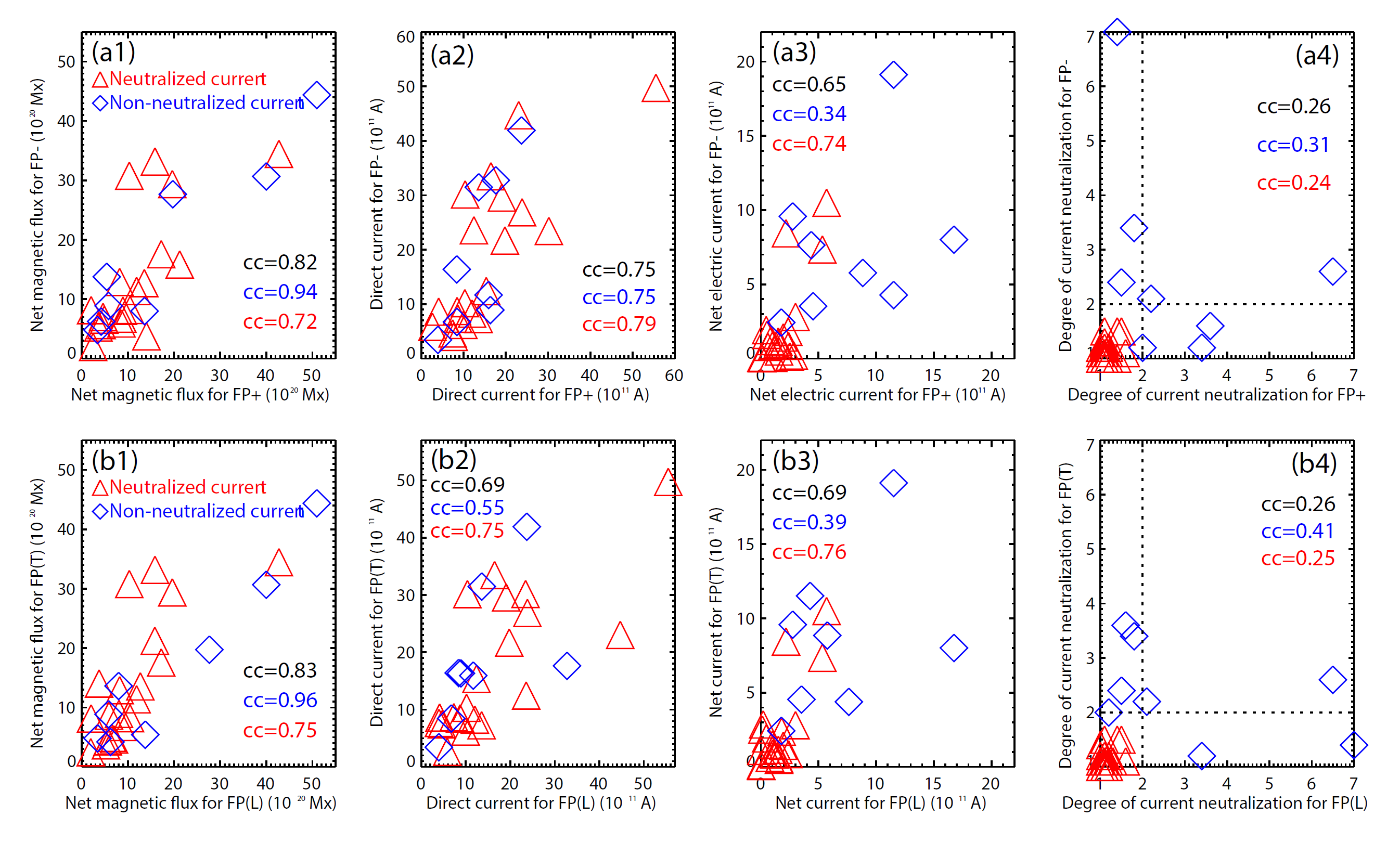}
	\caption{The correlation between magnetic properties measured in conjugate footpoints of studied MFRs. We compare net magnetic fluxes, direct electric currents, net electric currents and the degree of current neutralization for two footpoints. The red triangle is for the MFRs with neutralized current, while the blue rhombus is for the MFRs with non-neutralized current. The 'FP+/-' present the FP with positive/negative magnetic field, while the 'FP(L/T)' stand for the leading/trailing FP. The 'cc' is the absolute value of the cross correlation coefficient. \label{fig9}}
\end{figure}

\begin{figure}
	\plotone{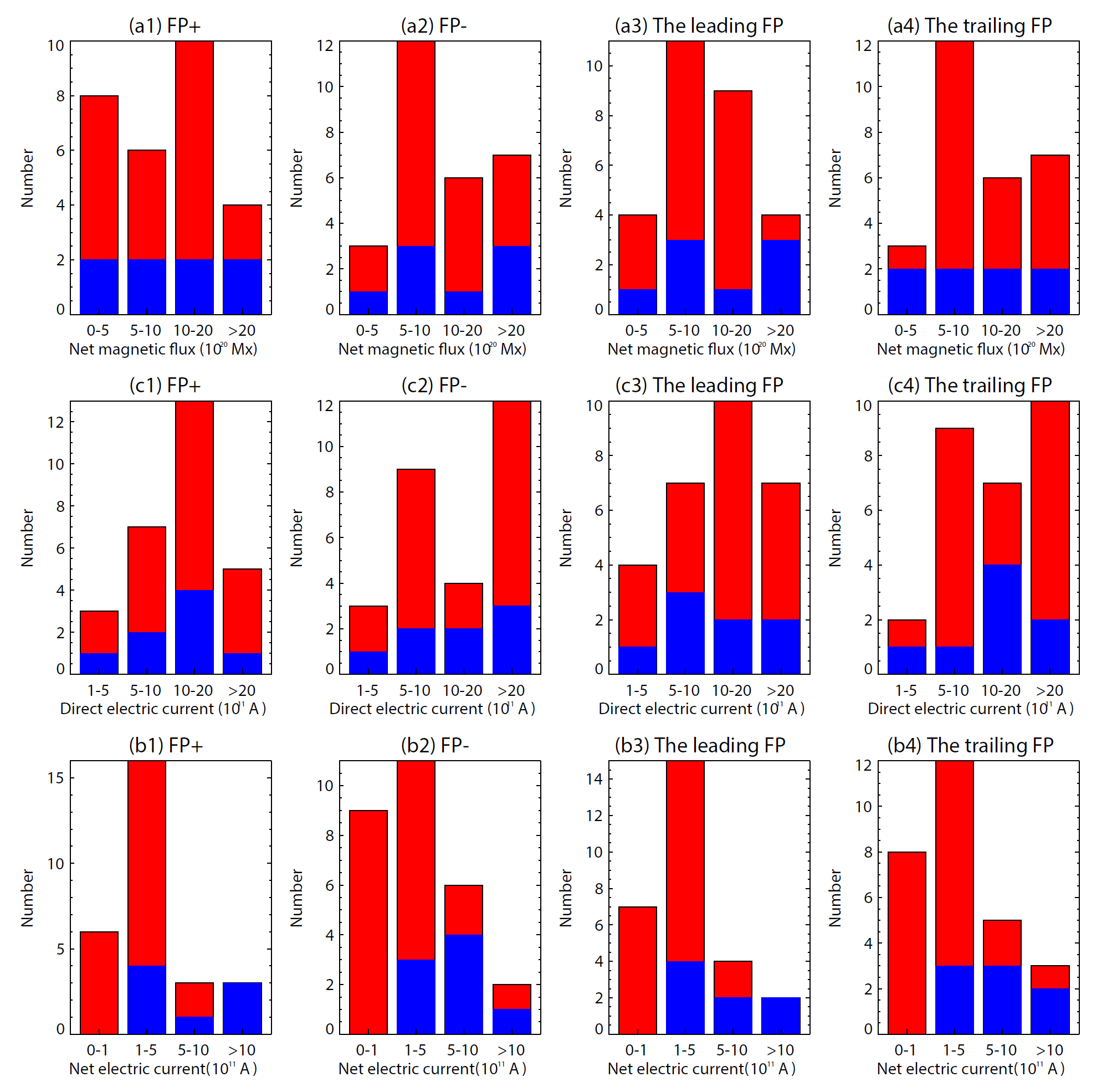}
	\caption{Comparison between properties of the MFRs with or without net currents. Red part represents the number distribution of the MFR with neutralized current, while blue part represents the other one. \label{fig10}}
\end{figure}

\begin{figure}
	\plotone{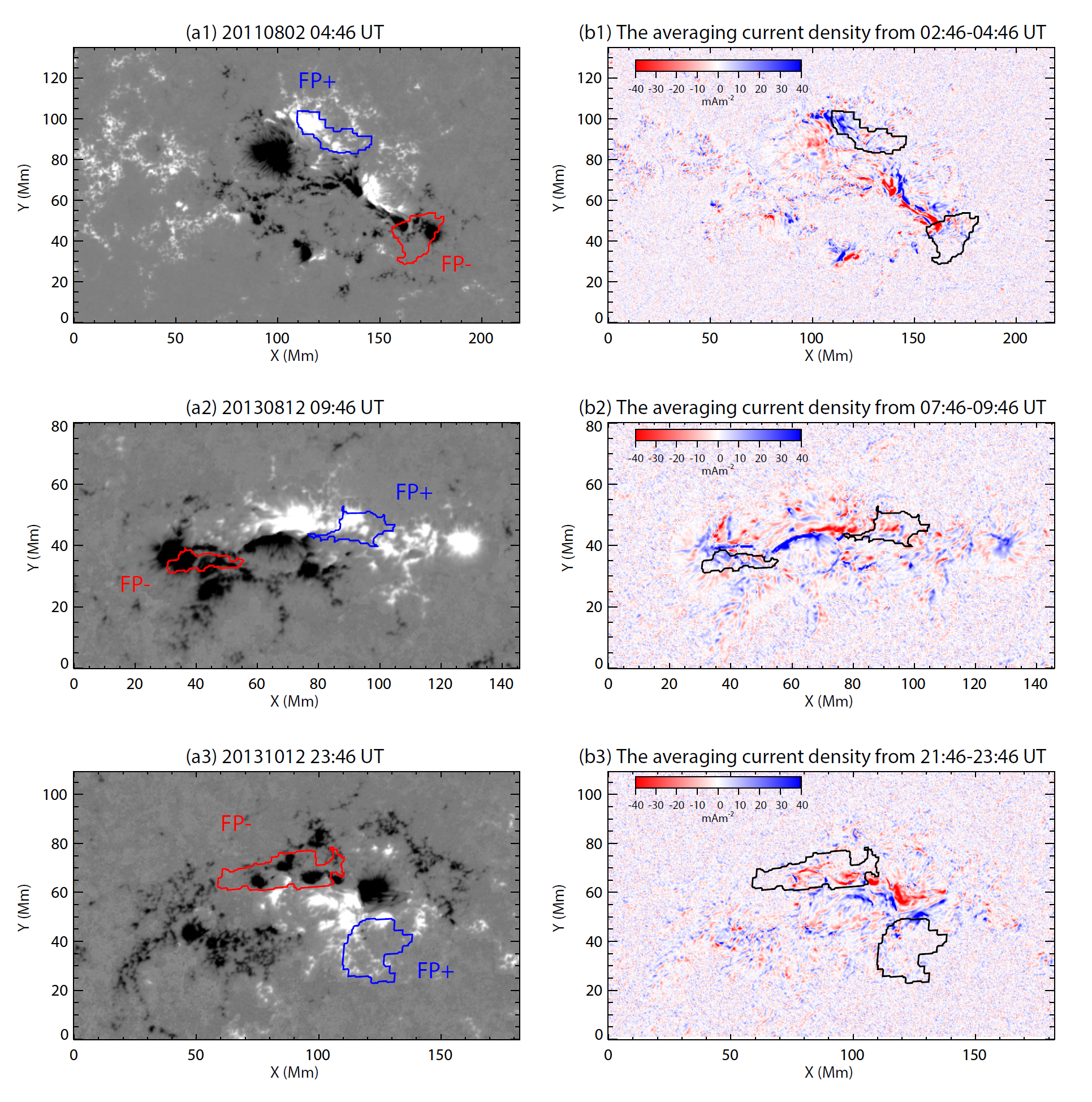}
	\caption{Three typical post-eruption dimmings with non-neutralized currents. The panels in the left are vertical magnetic field from HMI vector magnetogram. The panels in the right are the time-averaging current density maps. Two contours in each panel show the identified footpoint regions.\label{fig11}}
\end{figure}

\begin{figure}
	\plotone{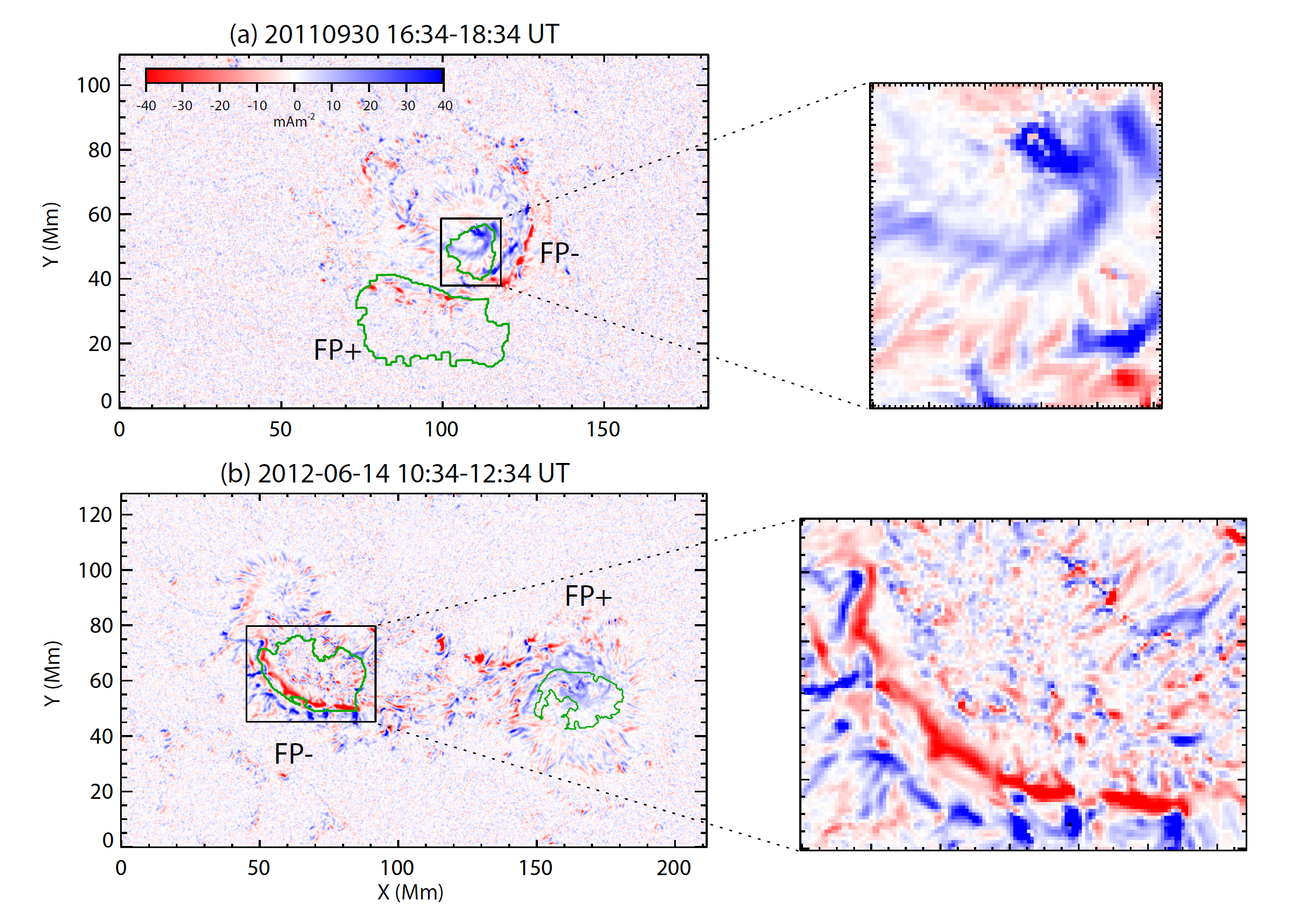}
	\caption{Asymmetric current distribution in the two feet of the MFRs. (a) and (b) are two snapshots of vertical current density maps for the 20110930 and 20120614 events. Two green contours in each panel represent the footpoints. \label{fig12}}
\end{figure}

\begin{figure}
	\plotone{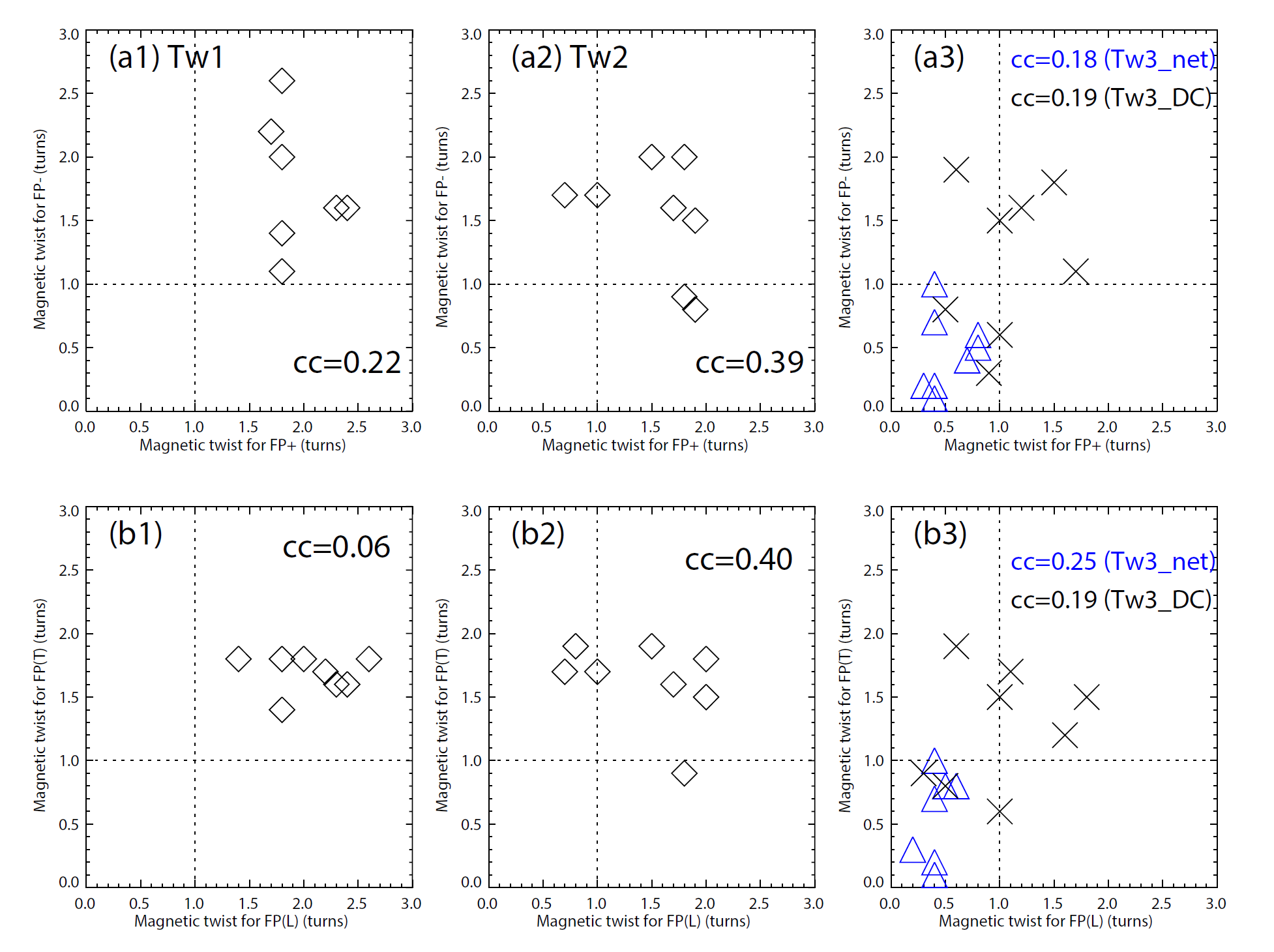}
	\caption{Comparison between magnetic twist in two footpoints of the MFRs with non-neutralized currents. Three columns show the average twist for three methods (Tw1,Tw2,Tw3) in two footpoints respectively. For the (a3) and (b3), the blue triangle is for $Tw3_{net}$, while the dark cross is for $Tw3_{DC}$. A critical value of magnetic twist (1.0 turn) are marked as two dashed lines in three panels. \label{fig13}}
\end{figure}

\clearpage
\appendix

\section{Identifying conjugate dimmings and associated uncertainties}\label{app:A}
Our previous studies \citep{wang2017buildup,wang2019evolution} had employed image thresholding segmentation methods to detect coronal dimmings. In short, the method would count all the pixels when their brightness are reduced by around 20\% to 40\% compared with their original value in the active region. However, morphological characteristics of coronal dimmings are complex, manifesting as multiple fragment-like areas or some large extended areas. The main problem is how to distinguish MFR feet-related dimmings from other type of dimmings. Previous studies (e.g \citealt{dissauer2018statistics}) considered so-called core dimmings as the feet-related dimmings. But the core dimmings were always shown as small fragments (see Figure 2 in the \cite{dissauer2018statistics}), even occurred at different locations. Inspired by 3D extension of flare/CME models (e.g. \citealt{moore2001onset,longcope2007quantitative,janvier2014electric}), for two-ribbon flares, feet-related dimmings should be co-spatial with flare ribbons. \cite{wang2017buildup} then took two closed hooks as boundaries of feet-related dimmings and finally attained two well identified footpoints of the MFR. Two studies \citep{qiu2017gradual,wang2019evolution} further confirmed that a pair of conjugate dimmings, which occur in the vicinity of flare ribbons and are associated with two opposite signs of magnetic polarities, map two footpoints of MFRs.
	
In this study, a series of steps mentioned in the Section~\ref{sec:fp} are used to seek two boundaries of conjugate dimmings automatically. Initially, for each event, our method will automatic check all images from the whole observational time and output two areas as the candidate conjugate dimming regions. Analyzing all EUV channels during several hours for each event will cost a lot of time. To minimize the amount of computation, we only check morphology of dimmings at several moments (e.g. before the eruption, the onset of flare, after the eruption) due to relatively slow evolution of dimming regions. In the following steps, all detected regions will be projected into the HMI vector magnetogram. The regions with mixed polarities will be deleted. Then we compare the location of flare ribbons and the detected dimming regions. After this processing, for most events, the remaining dimming regions will distribute mainly in two areas. But if the remaining dimming regions distribute in many different positions, we will further check magnetic connectivity of these positions via NLFFF models to further select two possible areas. In this study, only several events are required to check magnetic connectivity. Then the method will track the dimmings in AIA 304 channel at a cadence of 2 minutes starting from two hours before the eruption within two boundaries from previous steps. As explained in the Section~\ref{sec:fp}, two type of dimmings, pre-eruption and post-eruption, are observed in our study. For the pre-eruption dimmings, the method will keep going back in time to seek the onset of dimmings. By employing a trial-and-error approach, we find that the threshold of pre-eruption dimmings are around 15\% to 30\%. Similarly, we also recheck the evolution of post-eruption dimmings and find their threshold of around 30\% to 50\%. Figure~\ref{fig2} shows examples of dimming evolution.

The tracking of 28 events further show that the movement of dimmings is relatively complicated. As discussed in the Section~\ref{sec:dis-dim}, different dynamic evolution of dimmings are largely associated with the magnetic reconnection between the enveloped field and the MFR itself. Then the dimmed pixels only appear at the stage of rapid change are removing from our detection. We recheck the bright-curve of all detected pixels during the whole observational time. The flagged pixels with the brightness continues to drop at least 10 minutes are selected. Finally, we outline two regions, including all survived pixels, as two fixed regions.

It is very difficult to provide the uncertainties of identified footpoint regions. As mentioned above, a series of steps are employed to outline the footpoint regions. The errors of detected dimmings can be quantified by varying the threshold $\pm$ 10\% (see the errors in the Figure~\ref{fig2}). But locations and quantities of flagged pixels that are filtered through the aforementioned steps will be essentially unchanged when varying the threshold $\pm$ 10\%. The uncertainties of footpoint regions should depend on the initial boundaries of dimmings. As mentioned in the Section~\ref{sec:fp}, we analyze all EUV channels and the corresponding HMI data to acquire the boundaries. Unfortunately, it is very complicated to estimate the errors in these steps.

\section{Calculating electric current and associated uncertainties} \label{app:B}
Previous studies (e.g. \citealt{liu2017electric,avallone2020electric}) neglected the pixels with $|B|<200$ G when calculating the vertical electric currents from the HMI data to avoid regions with low signal-to-noise ratios. In this study, we have done a series of calculations to check how noise-like regions affect the estimation of electric current. Table~\ref{tab5} shows four results for electric current calculations. The original value of $I_{z}$ calculated by integrating all pixels in the footpoint regions is shown in the first column (labeled as (1) in the Table~\ref{tab5}). The second (2) and third (3) terms are two values of $I_{z}$ calculated by only considering the pixels with $|B|>100$ G or $|B|>200$ G respectively. The time-averaged value of $I_{z}$ is shown in the last column (labeled as (4) in the Table~\ref{tab5}. The results show that removing the weak-field pixels from calculations will reduce the value of $I_{z}$. For many events, if we neglect the pixels with $|B|<200$ G, the $I_{z}^{DC}$ will decrease by 2 to 4 times when comparing with the original value. The $R_{z}$ calculated from these different values of $I_{z}$ is listed in the last group of Table~\ref{tab5}. Removing weak-field pixels from calculations will increase the value of $R_{z}$. We find that the time-averaged value of $I_{z}$ is similar to the value of $I_{z}$ calculated by removing pixels with $|B|<100$ G, 0.5 to 1 times of the original value. But the time-averaged value of $R_{z}$ can better help us to distinct two MFR populations, with or without net current. We finally decide to take the time-averaging value of $I_{z}$ as our estimations in this study.

Here we also compare three different methods of estimating uncertainties of electric currents. 1) we follow the principle of the error propagation:
\[ \sqrt{\delta_{B_{x}(i,j+2)}^2 + \delta_{B_{x}(i,j+1)}^2 + \delta_{B_{x}(i,j-1)}^2 + \delta_{B_{x}(i,j-2)}^2+ \delta_{B_{y}(i+2,j)}^2 + \delta_{B_{y}(i+1,j)}^2 + \delta_{B_{y}(i-1,j)}^2 + \delta_{B_{y}(i-2,j)}^2}\]	
to attain the $\delta_{j_{z}}$ based on uncertainties from HMI data ($\delta_{Bx}$,$\delta_{By}$). 2) we tried a Monte Carlo simulation to calculate $\delta_{j_{z}}$ by randomly varying transverse field within uncertainties from HMI data. The simulation is conducted with $10^{5}$ iterations. Then the standard deviation of the Gaussian distribution of $j_{z}$ at each pixel is taken as $\delta_{j_{z}}$. 3) the standard derivation of the variation of $j_{z}$ during two hours prior to flare is taken as $\delta_{j_{z}}$. Table~\ref{tab6} shows three different errors for $I_{z}^{net}$, $I_{z}^{DC}$ and  $I_{z}^{RC}$. The errors calculated by the propagation equations are always larger than the another two methods. The smallest errors are always from the Monte Carlo simulation. There are no huge difference for three methods. As explained in the Section~\ref{sec:cur}, the time-averaged value of $j_{z}$ is selected to calculate electric current. It is suitable to take the standard derivation of $j_{z}$ during two hours as the errors here.

\begin{deluxetable*}{C|C|C|CCCC|CCCC|CCCC}[b!]
	\tabletypesize{\scriptsize}
	\tablecaption{Comparing the different value of electric currents\label{tab5}}
	\tablecolumns{16}
	\tablenum{5}
	\tablewidth{0pt}
	\tablehead{
		\colhead{} & \colhead{} & \colhead{} & \colhead{} & \colhead{} & \colhead{} & \colhead{} & \colhead{} & \colhead{} & \colhead{} & \colhead{} & \colhead{} & \colhead{} & \colhead{} & \colhead{}\\
		\colhead{No.} & \colhead{Date} & \colhead{} & \colhead{$I_{z}^{DC}$} & \colhead{($10^{11}$ A)} & \colhead{} & \colhead{} & \colhead{$I_{z}^{RC}$} & \colhead{($10^{11}$ A)} & \colhead{} & \colhead{} & \colhead{$R_{z}$} & \colhead{} & \colhead{} & \colhead{}\\
		\colhead{} & \colhead{} & \colhead{} &  \colhead{(1)} & \colhead{(2)} & \colhead{(3)} & \colhead{(4)} & \colhead{(1)} & \colhead{(2)} & \colhead{(3)} & \colhead{(4)} & \colhead{(1)} & \colhead{(2)} & \colhead{(3)} & \colhead{(4)}
	}
	\startdata
	1 & 20100807 & FP+ & 18.56  & -9.86 & -4.66 & -8.67 & -17.80 & 8.87 & 3.63 & 8.56 & 1.0 & 1.1 & 1.3 & 1.0\\
	  &          & FP- & -14.97 & 8.02 & 4.69 & -7.32 & 14.86 & -7.77 & -4.12 & 7.30 & 1.0 & 1.0 & 1.1 & 1.0 \\
	\hline
    2 & 20110307 & FP+ & 21.96  & -14.31 & -7.24 & -10.31 & 17.47 & 11.08 & 4.39 & 7.45 & 1.3 & 1.3 & 1.7 & 1.4\\
	  &          & FP- & -21.02 & 14.75 & 7.35 & 10.18 & 20.77 & -14.48 & -6.81 & -9.97 & 1.0 & 1.0 & 1.1 & 1.0 \\
	\hline
	3 & 20110621 & FP+ & 77.36  & 39.33 & 16.40 & 30.17 & -75.44 & -36.23 & -13.98 & -29.64 & 1.0 & 1.1 & 1.2 & 1.0\\
	  &          & FP- & -58.96 & -34.28 & 15.03 & -23.43 & 55.16 & 32.35 & -13.98 & 21.83 & 1.1 & 1.1 & 1.1 & 1.1 \\
	\hline
	4 & 20110802 & FP+ & 17.68  & 17.47 & 16.31 & 15.93 & -6.12 & -5.89 & -4.89 & -4.41 & 2.9 & 3.0 & 3.3 & 3.6\\
	  &          & FP- & -17.31 & -14.70 & -12.93 & -11.68 & 11.77 & 9.47 & 7.94 & 7.39 & 1.5 & 1.6 & 1.6 & 1.6 \\
	\hline
    5 & 20110930 & FP+ & -35.63  & -20.61 & -11.62 & -16.33 & 32.57 & 16.76 & 8.24 & 11.95 & 1.1 & 1.2 & 1.4 & 1.4\\
 	  &          & FP- & 9.40 & 9.40 & 9.40 & 8.92 & -2.10 & -2.10 & -2.10 & -1.27 & 4.5 & 4.5 & 4.5 & 7.0 \\
	\hline
	6 & 20120309 & FP+ & -36.17 & -29.02 & -19.80 & -19.77 & 30.24 & 23.47 & 14.48 & 14.43 & 1.2 & 1.2 & 1.4 & 1.4\\
	  &          & FP- & 34.46 & 33.08 & 28.15 & 21.69 & -26.23 & -25.75 & -21.37 & -14.34 & 1.3 & 1.3 & 1.3 & 1.5 \\
	\hline
	7 & 20120310 & FP+ & 49.74 & 35.16 & -19.38 & 23.84 & -46.98 & -34.10 & 17.71 & -21.65 & 1.1 & 1.0 & 1.1 & 1.1\\
	  &          & FP- & 44.37 & 41.86 & 34.95 & 26.87 & -35.96 & -33.22 & -26.87 & -18.43 & 1.2 & 1.3 & 1.3 & 1.5 \\
	\hline
	8 & 20120614 & FP+ & 14.38 & 14.38 & 14.38 & 13.60 & -3.57 & -3.57 & -3.57 & -2.08 & 4.0 & 4.0 & 4.0 & 6.5\\
	  &          & FP- & -39.67 & -39.27 & -37.95 & -31.47 & 20.03 & 19.84 & 18.98 & 12.35 & 2.0 & 2.0 & 2.0 & 2.6 \\
	\hline
    9 & 20120712 & FP+ & 112.72 & 72.42 & 41.41 & 55.35 & -110.36 & -71.49 & -40.28 & -53.06 & 1.0 & 1.0 & 1.0 & 1.0 \\
	&          & FP- & 104.67 & 57.94 & 34.24 & -49.94 & -104.53 & -56.88 & -32.16 & 48.87 & 1.0 & 1.0 & 1.1 & 1.0 \\
	\hline
	10 & 20130206 & FP+ & 33.18 & 26.34 & 12.24 & 15.36 & -30.84 & -23.32 & -9.22 & -12.98 & 1.0 & 1.1 & 1.3 & 1.2\\
	&          & FP- & -30.41 & -21.26 & -7.04 & -12.40 & 28.96 & 20.15 & 5.95 & 10.59 & 1.1 & 1.1 & 1.2 & 1.2 \\
	\hline
    11 & 20130411 & FP+ & -76.29 & -39.06 & 9.29 & -25.94 & 75.27 & 37.57 & -8.63 & 25.33 & 1.0 & 1.0 & 1.1 & 1.0\\
	  &          & FP- & 9.76 & 8.87 & 6.81 & 5.89 & -9.18 & -8.13 & -5.82 & -5.03 & 1.1 & 1.1 & 1.2 & 1.2 \\
	\hline
	12 & 20130517 & FP+ & -15.90 & -13.71 & -10.29 & -10.37 & 13.94 & 11.99 & 8.36 & 8.42 & 1.1 & 1.1 & 1.2 & 1.2\\
	&          & FP- & 67.00 & 26.12 & 9.05 & 30.16 & -65.34 & -25.71 & -8.98 & -28.74 & 1.0 & 1.0 & 1.0 & 1.1 \\
	\hline
	13 & 20130812 & FP+ & -4.88 & -4.74 & -4.37 & -3.95 & 3.56 & 3.17 & 2.59 & 2.17 & 1.4 & 1.5 & 1.7 & 1.8\\
	&          & FP- & 3.79 & 3.79 & 3.77 & 3.43 & -2.50 & -2.50 & -2.49 & -1.01 & 1.5 & 1.5 & 1.5 & 3.4 \\
	\hline
    14 & 20130817 & FP+ & 7.44 & 5.82 & 3.77 & 4.12 & -5.85 & -4.50 & -2.43 & -2.53 & 1.3 & 1.3 & 1.6 & 1.6\\
	   &          & FP- & 18.00 & 14.82 & -8.30 & -8.61 & -17.92 & -14.62 & 8.07 & 8.14 & 1.0 & 1.0 & 1.0 & 1.1 \\
	\hline
	15 & 20130830 & FP+ & 50.47 & 35.31 & 11.95 & 21.85 & -48.94 & -32.72 & -10.49 & -19.66 & 1.0 & 1.1 & 1.1 & 1.1\\
	&          & FP- & -121.97 & 52.64 & -10.25 & 45.46 & 120.12 & -52.37 & 10.02 & -45.14 & 1.0 & 1.0 & 1.0 & 1.0 \\
	\hline
    16 & 20131013 & FP+ & 14.10 & 11.66 & 7.63 & 8.40 & -10.86 & -8.02 & -4.62 & -5.63 & 1.3 & 1.5 & 1.7 & 1.5\\
	&          & FP- & -24.60 & -22.68 & -18.80 & -16.39 & 13.98 & 12.46 & 9.68 & 6.81 & 1.8 & 1.8 & 1.9 & 2.4 \\
	\hline
	17 & 20140131 & FP+ & -22.65 & -16.04 & 3.35 & -8.47 & 22.25 & 15.24 & -2.75 & 8.41 & 1.0 & 1.1 & 1.2 & 1.0\\
	&          & FP- & 20.17 & -14.95 & -5.28 & 8.73 & -18.82 & 14.57 & 3.51 & -8.67 & 1.1 & 1.0 & 1.5 & 1.0 \\
	\hline
    18 & 20140320 & FP+ & 18.81 & 15.25 & 8.21 & 9.96 & -18.46 & -13.75 & -6.47 & -9.49 & 1.0 & 1.1 & 1.3 & 1.1\\
	&          & FP- & -9.06 & -8.47 & -7.18 & -6.18 & 7.12 & 6.55 & 5.45 & 4.26 & 1.3 & 1.3 & 1.3 & 1.5 \\
	\hline
	19 & 20140730 & FP+ & -21.33 & -19.61 & -15.39 & -12.04 & 20.46 & 19.11 & 13.69 & 11.02 & 1.0 & 1.0 & 1.1 & 1.1\\
	&          & FP- & 16.39 & -13.09 & -6.89 & 8.25 & -15.78 & 12.78 & 6.01 & -7.36 & 1.0 & 1.0 & 1.2 & 1.1 \\
	\hline
    20 & 20140801 & FP+ & 25.15 & 16.25 & 11.87 & 13.56 & -23.40 & -14.37 & -9.78 & -12.04 & 1.1 & 1.1 & 1.2 & 1.1\\
	&          & FP- & 13.40 & 10.67 & 7.32 & 7.27 & -12.78 & -10.13 & -6.58 & 6.12 & 1.1 & 1.1 & 1.1 & 1.2 \\
	\hline
	21 & 20140825 & FP+ & -16.08 & -12.02 & -8.44 & -8.45 & 11.27 & 7.72 & 4.57 & 3.91 & 1.4 & 1.6 & 1.9 & 2.2\\
	&          & FP- & 7.66 & 7.62 & 7.33 & 6.74 & -4.75 & -4.63 & -4.09 & -3.21 & 1.6 & 1.7 & 1.8 & 2.1 \\
	\hline
    22 & 20140825 & FP+ & -21.56 & -12.27 & -3.51 & 7.41 & 20.50 & 11.35 & 2.84 & 6.58 & 1.1 & 1.1 & 1.2 & 1.1\\
	&          & FP- & 7.35 & 6.11 & 4.79 & 4.62 & -6.38 & -5.08 & -3.14 & -3.44 & 1.2 & 1.2 & 1.5 & 1.3 \\
	\hline
	23 & 20140908 & FP+ & -26.99 & -26.99 & -26.94 & -23.71 & 9.53 & 9.53 & 9.44 & 6.96 & 2.8 & 2.8 & 2.9 & 3.4\\
	&          & FP- & 67.42 & 53.86 & 36.25 & 41.94 & -59.67 & -47.54 & -31.48 & -33.93 & 1.1 & 1.1 & 1.2 & 1.2 \\
	\hline
    24 & 20140910 & FP+ & -21.46 & -21.38 & -20.16 & -16.47 & 15.64 & 15.40 & 14.24 & 10.76 & 1.4 & 1.4 & 1.4 & 1.5\\
	&          & FP- & 50.05 & 43.95 & 36.02 & 33.48 & -41.29 & -35.50 & -26.56 & -22.99 & 1.2 & 1.2 & 1.4 & 1.5 \\
	\hline
	25 & 20140921 & FP+ & 16.46 & 8.56 & -3.61 & 7.67 & -15.93 & -8.45 & 3.58 & -6.59 & 1.0 & 1.0 & 1.0 & 1.2\\
	&          & FP- & 6.94 & 5.92 & 3.48 & 3.98 & -6.90 & -5.84 & -3.32 & -3.93 & 1.0 & 1.0 & 1.0 & 1.0 \\
	\hline
	26 & 20141220 & FP+ & 30.59 & 30.42 & 28.79 & 19.08 & -27.50 & -27.21 & -25.46 & -16.08 & 1.1 & 1.1 & 1.1 & 1.2\\
	&          & FP- & -45.85 & -43.18 & -33.16 & -29.58 & 42.42 & 39.63 & 31.75 & 26.77 & 1.1 & 1.1 & 1.0 & 1.1 \\
	\hline
	27 & 20150622 & FP+ & -21.35 & -20.98 & -19.85 & -17.64 & 12.97 & 12.41 & 10.75 & 8.79 & 1.7 & 1.7 & 1.9 & 2.0\\
	&          & FP- & 64.19 & 48.36 & 33.26 & 32.75 & -58.04 & -42.25 & -26.12 & -26.98 & 1.1 & 1.2 & 1.3 & 1.2 \\
	\hline
	28 & 20151104 & FP+ & 24.16 & 22.80 & 19.49 & 12.45 & -23.16 & -22.12 & -17.97 & -10.97 & 1.0 & 1.0 & 1.0 & 1.1 \\
	&          & FP- & 54.76 & -49.27 & -24.54 & -23.57 & -54.29 & 48.72 & 23.16 & 22.55 & 1.0 & 1.0 & 1.0 & 1.1 \\
	\hline
	\enddata
	\tablecomments{ }
\end{deluxetable*}

\begin{deluxetable*}{C|C|C|CCC|CCC|CCC}[b!]
	\tabletypesize{\scriptsize}
	\tablecaption{Errors for direct currents inside the 28 MFRs\label{tab6}}
	\tablecolumns{12}
	\tablenum{6}
	\tablewidth{0pt}
	\tablehead{
		\colhead{No.} & \colhead{Date} & \colhead{} & \colhead{$\delta_{I_{z}^{net}}$} & \colhead{($10^{11}$ A)} & \colhead{} & \colhead{$\delta_{I_{z}^{DC}}$} & \colhead{($10^{11}$ A)} & \colhead{} & \colhead{$\delta_{I_{z}^{RC}}$} & \colhead{($10^{11}$ A)} & \colhead{}\\
		\colhead{} & \colhead{} & \colhead{} &  \colhead{EP} & \colhead{MCS} & \colhead{SD} & \colhead{EP} & \colhead{MCS} & \colhead{SD} & \colhead{EP} & \colhead{MCS} & \colhead{SD}
	}
	\startdata
	1 & 20100807 & FP+ & 1.88 & 0.72 & 1.31 & 1.40 & 0.61 & 0.64 & 1.27 & 0.37 & 0.65 \\
	  &          & FP- & 1.66 & 0.64 & 1.15 & 1.04 & 0.63 & 0.54 & 1.04 & 0.47 & 0.50\\
	  \hline
	2 & 20110307 & FP+ & 1.85 & 0.82 & 1.31 & 1.23 & 0.69 & 0.65 & 1.34 & 0.45 & 0.64\\
	  &          & FP- & 1.96 & 0.78 & 1.29 & 1.46 & 0.63 & 0.62 & 1.44 & 0.47 & 0.60\\
	  \hline
	3 & 20110621 & FP+ & 3.16 & 1.48 & 1.82 & 3.77 & 1.06 & 1.26 & 3.73 & 1.03 & 1.24\\
      &          & FP- & 3.10 & 1.29 & 1.68 & 3.66 & 0.93 & 1.07 & 3.61 & 0.90 & 1.06\\
      \hline
	4 & 20110802 & FP+ & 1.66 & 0.73 & 0.91 & 1.09 & 0.66 & 0.36 & 0.99 & 0.32& 0.27\\
      &          & FP- & 1.86 & 0.62 & 1.16 & 1.41 & 0.59 & 0.64 & 1.30 & 0.18 & 0.50\\
      \hline
	5 & 20110930 & FP+ & 2.33 & 1.09 & 1.49 & 2.11 & 1.00 & 0.91 & 1.98 & 0.52 & 0.78\\
      &          & FP- & 2.61 & 0.62 & 1.06 & 2.57 & 0.58 & 0.42 & 2.02 & 0.18 & 0.21\\
      \hline
	6 & 20120309 & FP+ & 2.64 & 1.02 & 1.48 & 2.72 & 0.85 & 0.82 & 2.63 & 0.57 & 0.80\\
      &          & FP- & 2.72 & 1.14 & 1.45 & 2.89 & 0.83 & 0.87 & 2.70 & 0.77 & 0.79\\
      \hline  
    7 & 20120310 & FP+ & 2.50 & 1.30 & 1.61 & 2.36 & 0.81 & 0.97 & 2.32 & 1.02 & 0.91\\
      &          & FP- & 2.61 & 1.24 & 1.46 & 2.54 & 0.87 & 0.82 & 2.61 & 0.87 & 0.78 \\
      \hline 
    8 & 20120614 & FP+ & 1.90 & 0.51 & 0.87 & 1.37 & 0.47 & 0.28 & 0.78 & 0.20 & 0.15\\
      &          & FP- & 2.06 & 1.33 & 1.32 & 1.72 & 1.16 & 0.70 & 1.49 & 0.64 & 0.61 \\
      \hline 
    9 & 20120712 & FP+ & 2.81 & 1.73 & 1.87 & 2.97 & 1.02 & 1.34 & 3.23 & 1.40 & 1.32\\
      &          & FP- & 2.91 & 1.68 & 1.84 & 3.15 & 1.00 & 1.26 & 2.90 & 1.36 & 1.28 \\
      \hline 
    10& 20130206 & FP+ & 2.35 & 1.01 & 1.41 & 2.11 & 0.62 & 0.73 & 2.06 & 0.80 & 0.76 \\
      &          & FP- & 2.17 & 0.99 & 1.42 & 1.81 & 0.81 & 0.78 & 1.77 & 0.56 & 0.76\\
      \hline 
    11& 20130411 & FP+ & 2.98 & 1.60 & 1.80 & 3.40 & 1.33 & 1.23 & 3.32 & 0.88 & 1.21 \\
      &          & FP- & 1.55 & 0.53 & 0.93 & 0.91 & 0.36 & 0.32 & 0.82 & 0.59 & 0.36\\
      \hline
    12& 20130517 & FP+ & 1.72 & 0.64 & 1.09 & 1.13 & 0.52 & 0.45 & 1.12 & 0.36 & 0.45 \\
      &          & FP- & 2.70 & 1.33 & 1.78 & 2.79 & 0.62 & 1.19 & 2.70 & 1.17 & 1.19\\
      \hline
    13& 20130812 & FP+ & 1.23 & 0.43 & 0.79 & 0.60 & 0.35 & 0.24 & 0.57 & 0.25 & 0.23 \\
      &          & FP- & 1.18 & 0.40 & 0.86 & 0.54 & 0.30 & 0.32 & 0.50 & 0.26 & 0.24 \\
      \hline
    14& 20130817 & FP+ & 1.45 & 0.47 & 0.93 & 0.80 & 0.32 & 0.30 & 0.74 & 0.34 & 0.33\\
      &          & FP- & 2.07 & 0.75 & 1.19 & 1.59 & 0.47 & 0.54 & 1.63 & 0.59 & 0.53\\
      \hline
    15& 20130830 & FP+ & 3.05 & 1.16 & 1.61 & 3.55 & 0.62 & 0.99 & 4.55 & 0.98 & 0.96\\
      &          & FP- & 3.48 & 1.92 & 2.08 & 4.57 & 1.67 & 1.64 & 3.44 & 0.95 & 1.64\\
      \hline
    16& 20131013 & FP+ & 1.79 & 0.62 & 1.14 & 1.21 & 0.43 & 0.49 & 1.10 & 0.44 & 0.47\\
      &          & FP- & 1.87 & 1.07 & 1.23 & 1.39 & 0.88 & 0.63 & 1.25 & 0.60 & 0.51 \\
      \hline
    17& 20140131 & FP+ & 2.10 & 0.95 & 1.44 & 1.68 & 0.77 & 0.80 & 1.63 & 0.55 & 0.76 \\
      &          & FP- & 2.09 & 0.87 & 1.39 & 1.64 & 0.55 & 0.73 & 1.59 & 0.68 & 0.70 \\
      \hline
    18& 20140320 & FP+ & 2.03 & 0.75 & 1.27 & 1.55 & 0.48 & 0.60 & 1.54 & 0.58 & 0.61 \\
      &          & FP- & 1.41 & 0.55 & 0.93 & 0.77 & 0.43 & 0.35 & 0.73 & 0.34 & 0.31 \\
      \hline
    19& 20140730 & FP+ & 1.10 & 0.50 & 1.18 & 2.00 & 1.00 & 0.56 & 1.10 & 0.50 & 0.50 \\
      &          & FP- & 1.40 & 0.40 & 1.31 & 1.20 & 0.40 & 0.65 & 1.30 & 0.30 & 0.70\\
      \hline
    20& 20140801 & FP+ & 2.06 & 0.93 & 1.30 & 1.60 & 0.58 & 0.46 & 1.56 & 0.73 & 0.67 \\
      &          & FP- & 1.84 & 0.61 & 1.13 & 1.25 & 0.39 & 0.50 & 1.29 & 0.48 & 0.63\\
      \hline
    21& 20140825 & FP+ & 1.62 & 0.77 & 1.04 & 1.02 & 0.67 & 0.45 & 0.95 & 0.37 & 0.37 \\
      &          & FP- & 1.25 & 0.57 & 0.82 & 0.59 & 0.47 & 0.26 & 0.45 & 0.32 & 0.26 \\
      \hline
    22& 20140825 & FP+ & 1.70 & 0.79 & 1.19 & 1.09 & 0.58 & 0.55 & 1.08 & 0.40 & 0.52 \\
      &          & FP- & 1.20 & 0.50 & 0.84 & 0.54 & 0.36 & 0.27 & 0.56 & 0.34 & 0.28 \\
      \hline
    23& 20140908 & FP+ & 2.17 & 0.82 & 1.01 & 5.47 & 0.70 & 0.44 & 1.78 & 0.40 & 0.38 \\
      &          & FP- & 2.57 & 1.39 & 1.69 & 2.53 & 0.94 & 1.13 & 2.47 & 1.02 & 1.02 \\
      \hline
    24& 20140910 & FP+ & 1.85 & 0.89 & 1.23 & 1.37 & 0.65 & 0.50 & 1.29 & 0.61 & 0.57 \\
      &          & FP- & 2.85 & 1.28 & 1.55 & 3.15 & 0.95 & 0.95 & 2.99 & 0.87 & 0.87 \\
      \hline
    25& 20140921 & FP+ & 1.72 & 0.65 & 1.19 & 1.10 & 0.46 & 0.55 & 1.12 & 0.45 & 0.52 \\
      &          & FP- & 1.35 & 0.38 & 0.89 & 0.69 & 0.26 & 0.30 & 0.69 & 0.27 & 0.29\\
      \hline
    26& 20141220 & FP+ & 2.22 & 1.80 & 1.55 & 1.86 & 0.80 & 0.90 & 4.23 & 0.80 & 0.86 \\
      &          & FP- & 2.38 & 0.70 & 1.42 & 2.15 & 1.50 & 0.77 & 2.10 & 1.20 & 0.76 \\
      \hline
    27& 20150622 & FP+ & 1.77 & 0.81 & 1.04 & 1.22 & 0.69 & 0.46 & 1.11 & 0.43 & 0.35 \\
      &          & FP- & 3.10 & 1.44 & 1.69 & 3.34 & 1.08 & 1.02 & 3.61 & 0.96 & 1.08 \\
      \hline
    28& 20151104 & FP+ & 2.52 & 0.50 & 1.33 & 2.30 & 0.80 & 0.67 & 2.49 & 0.50 & 0.66\\
      &          & FP- & 2.53 & 1.00 & 1.88 & 2.41 & 0.70 & 1.33 & 2.36 & 0.50 & 1.36\\
      \hline
    \enddata
	\tablecomments{ }
\end{deluxetable*}

\end{document}